\documentclass[10pt,a4paper,twocolumn]{article}

\usepackage[a4paper,left=1.8cm,right=1.8cm,top=2.2cm,bottom=2.4cm,columnsep=0.7cm]{geometry}

\usepackage{graphicx}%
\usepackage{multirow}%
\usepackage{amsmath,amssymb,amsfonts}%
\usepackage{amsthm}%
\usepackage{mathrsfs}%
\usepackage[title]{appendix}%
\usepackage[dvipsnames,table]{xcolor}%
\usepackage{textcomp}%
\usepackage{manyfoot}%
\usepackage{booktabs}%
\usepackage{algorithm}%
\usepackage{mhchem} 
\usepackage{algorithmicx}%
\usepackage{algpseudocode}%
\usepackage{listings}%
\usepackage{tikz}
\usepackage{comment}
\usepackage{threeparttable}
\graphicspath{{./Figures/}}

\definecolor{DarkBlackGreen}{RGB}{30,80,00}

\usepackage{dblfloatfix}
\usepackage{array}

\usepackage{fancyhdr}
\usepackage{titlesec}
\usepackage{cuted}

\usepackage[
  font=small,
  labelfont=bf,
  labelsep=period,
  justification=justified
]{caption}
\usepackage{subcaption}

\usepackage{authblk}
\usepackage{cite}
\usepackage[colorlinks=true,linkcolor=RoyalBlue,citecolor=RoyalBlue,urlcolor=RoyalBlue,breaklinks=true]{hyperref}
\usepackage[switch]{lineno}

\theoremstyle{plain}%
\newtheorem{theorem}{Theorem}%
\newtheorem{proposition}[theorem]{Proposition}%
\theoremstyle{definition}%

\titleformat{\section}{\normalfont\large\bfseries}{\thesection.}{0.5em}{}
\titleformat{\subsection}{\normalfont\normalsize\bfseries}{\thesubsection.}{0.5em}{}
\titleformat{\subsubsection}{\normalfont\normalsize\itshape}{\thesubsubsection.}{0.5em}{}
\titlespacing*{\section}{0pt}{1.3em plus 0.25em minus 0.2em}{0.7em}
\titlespacing*{\subsection}{0pt}{1.0em plus 0.1em minus 0.2em}{0.5em}
\titlespacing*{\subsubsection}{0pt}{0.8em plus 0.1em minus 0.2em}{0.4em}

\fancypagestyle{firstpage}{%
  \fancyhf{}%
  \fancyfoot[C]{\small\thepage}%
}
\usepackage{cleveref}

\crefname{figure}{Fig.}{Figs.}
\Crefname{figure}{Fig.}{Figs.}
\crefname{section}{Sec.}{Secs.}
\Crefname{section}{Sec.}{Secs.}
\crefname{subsection}{Sec.}{Secs.}
\Crefname{subsection}{Sec.}{Secs.}
\crefname{appendix}{Appendix}{Appendices}
\Crefname{appendix}{Appendix}{Appendices}
\crefname{subappendix}{Appendix}{Appendices}
\Crefname{subappendix}{Appendix}{Appendices}
\crefname{equation}{Eq.}{Eqs.}
\Crefname{equation}{Eq.}{Eqs.}

\crefname{table}{Tab.}{Tabs.}
\Crefname{table}{Tab.}{Tabs.}

\newcommand{\Ron}{R_{\mathrm{on}}}
\newcommand{\Roff}{R_{\mathrm{off}}}
\newcommand{\Rth}{R_{\mathrm{th}}}
\newcommand{\Cth}{C_{\mathrm{th}}}
\newcommand{\Tamb}{T_{\mathrm{amb}}}
\newcommand{\Ea}{E_{\mathrm{a}}}
\newcommand{\kb}{k_{\mathrm{B}}}
\newcommand{\muv}{\mu_{\mathrm{v}}}

\usepackage{balance}    
\usepackage{etoolbox}
\usepackage{enumitem}   
\newif\ifrevision
\revisionfalse
\ifrevision
  \definecolor{newblue}{RGB}{0,72,196}
  \newcommand{\new}[1]{\textcolor{newblue}{#1}}
  \newenvironment{newpar}{\par\color{newblue}}{\par}
\else
  \newcommand{\new}[1]{#1}
  
\fi

\begin{document}

\twocolumn[
\begin{center}
{\LARGE\bfseries Electrothermal Compact Drift Model of TiO$_2$ Memristors:\\ Verilog-A Implementation and Dimensionless Regime Mapping\par}
\vspace{0.9em}
{\large
N.G.~Koudafok\^e$^{1,2,\ast}$,\;
Hilda~A.~Cerdeira$^{1}$,\;
L.~A.~Hinvi$^{3}$,\;
A.~V.~Monwanou$^{2}$
\par}

\vspace{0.6em}

{\small
$^{1}$ICTP South American Institute for Fundamental Research,
Instituto de F\'{i}sica Te\'{o}rica (IFT--UNESP), Bloco~II,
Rua Dr.\ Bento Teobaldo Ferraz 271, Barra Funda,
S\~ao Paulo, 01140-070, Brazil\\

$^{2}$Institut de Math\'{e}matiques et de Sciences Physiques (IMSP),
Universit\'e d'Abomey Calavi (UAC),
Porto-Novo, B\'enin\\

$^{3}$Laboratoire des Proc\'{e}d\'{e}s et de l'Innovation Technologiques (LaPIT), Institut National Sup\'erieur de Technologie Industrielle (INSTI) Lokossa/UNSTIM, Abomey, B\'enin 

$^{\ast}$Corresponding author:
\texttt{gilles.koudafoke@unesp.br}

\par}
\end{center}
\vspace{0.8em}
{\rule{\textwidth}{0.6pt}}
\vspace{0.4em}

\noindent{\bfseries Abstract.\ }%
Compact models of titanium-dioxide memristors used in circuit simulation
commonly follow the drift formulation of Strukov \textit{et al.} and assume
a constant ionic mobility, although oxygen-vacancy migration is thermally
activated and Joule self-heating is unavoidable. We present ATDM (Arrhenius
Thermally Activated Drift Model), a minimal electrothermal compact model
that couples the drift equation to a lumped heat balance through an
Arrhenius mobility. It adds one thermal state while preserving the
electrical state variable, the relation $v=iR(x)$, and the isothermal
limit. The model is implemented in \textsc{Verilog-A}, compiled with
OpenVAF, and simulated as a device in \texttt{ngspice}, where it
reproduces an independent reference implementation within $0.009\,\%$
of the peak voltage and recovers the isothermal excursion at zero
activation energy. A dimensionless formulation introduces a thermal lag
$\varepsilon$ and an effective switching number $\Theta$. Across
$2.5\times10^{5}$ simulations, $\Theta$ orders the state-variable
excursion over the tested quasi-static domain with a robust relative
scatter of $4\,\%$ and no fitted constant, and the $(\varepsilon,\Theta)$
regime map quantifies when the quasi-static thermal reduction remains
valid. Under current drive, the stroboscopic map of the reduced model is
proved to be strictly increasing, which excludes period-doubling and
chaos in that reduction. Within the sampled ranges, a variance-based
sensitivity analysis identifies the effective thermal resistance as a
first-order contributor to self-heating, while separate capacitance sweeps
show a weak influence of the thermal capacitance in the quasi-static regime. The results characterize the specified model with representative, uncalibrated parameters.

\vspace{0.5em}

\vspace{0.4em}
{\rule{\textwidth}{0.6pt}}
\vspace{1.2em}
]
\thispagestyle{firstpage}

\section{Introduction}
\label{sec:intro}
Since Chua's prediction of the memristor \cite{Chua1971}, its
generalization to memristive systems \cite{Chua1976}, and the
titanium-dioxide device reported by Strukov \textit{et al.}
\cite{Strukov2008}, memristive devices have attracted sustained
interest for nonvolatile memory, neuromorphic computing, and
reconfigurable circuits
\cite{Yang2013,Waser2007,Pershin2011,DiVentra2009}.
Switching in \ce{TiO2} is commonly associated with the field-driven
redistribution of oxygen vacancies
\cite{Yang2008,Szot2006,Jeong2009,Miao2011}. The Strukov model
represents this process through a single normalized state variable,
the relative length of a vacancy-rich region, governed by a linear
current-driven drift equation. This coarse-grained description,
retained here, does not resolve a conductive filament or its
geometry. Its ability to reproduce pinched hysteresis with few
parameters and low computational cost has made it a widely used
reference for window-function refinements
\cite{Joglekar2009,Biolek2009,Prodromakis2011} and more general
compact formulations, including TEAM and VTEAM
\cite{Kvatinsky2013,Kvatinsky2015}. The original model is,
however, isothermal: ionic mobility is constant, and temperature
is absent as a dynamical state.

This assumption suppresses a physically relevant feedback.
Vacancy migration is thermally activated, with computed barriers
of approximately $0.2$--$0.8$~eV depending on crystallographic
environment and vacancy charge state
\cite{Janotti2010,Morgan2010,DeSouza2015,Morgan2009,Schaub2003,DeLile2023}.
Joule heating can therefore alter the mobility governing
switching, while the evolving electrical state changes resistance
and hence the dissipated power. Thermal effects are reported to
influence switching speed, retention, and endurance in oxide
devices \cite{Ielmini2011,Valov2011,Kim2012,Mickel2014}.
Singh and Raj \cite{Singh2018} developed an analytical treatment
of temperature-dependent drift, and thermosensitive changes of
dynamical regime have also been reported in a memristive
\textsc{mems} model \cite{Koudafoke2026}. These studies motivate
the question addressed here: how does closing the electrothermal
feedback loop change the response of the Strukov model while
preserving its original state description and constitutive
relation?

For circuit design, two distinct criteria must be separated.
\new{The feedback strength $\Lambda$ of \cref{eq:Lambda}, which measures
the Arrhenius mobility enhancement produced by self-heating, determines
whether that correction is negligible and an isothermal description
suffices.}
Thermal lag determines whether temperature can be
represented algebraically or requires its own dynamical state.
The latter depends on the timescale of the dissipated power,
which need not coincide with that of the applied excitation.
\new{As general motivation, memristive devices operate over a broad
range of timescales: switching with $105$--$120$~ps pulses has been
measured in a tantalum oxide device \cite{Torrezan2011}, a material
system distinct from the one modeled here, and exponential
ionic-drift models provide conditions for nanosecond-scale
switching \cite{StrukovWilliams2009}.}
The present analysis expresses the feedback strength and thermal
lag through dimensionless quantities and quantifies the resulting
operating regimes
(\cref{sec:regimemap,sec:qs-frequency-range}).

We introduce ATDM (Arrhenius Thermally Activated Drift Model),
a minimal extension coupling the Strukov drift equation to a
lumped heat balance through an Arrhenius mobility. Electrothermal
coupling is already established in compact modeling, including
filamentary descriptions
\cite{Pickett2009,Kim2012,Mickel2014}; the contribution here is
to formulate and analyze it under the constraints of the Strukov
structure. ATDM preserves the original electrical state variable
and the relation $v=iR(x)$, adds only one temperature state,
and introduces neither filament geometry nor an imposed
switching threshold. It recovers the isothermal equation
analytically, with numerical agreement checked against an
independent reference implementation, and admits a controlled
timescale-separation reduction back to one state variable.
A \textsc{Verilog-A} implementation, compiled with OpenVAF
and executed as a device in \texttt{ngspice}, makes the model
\new{usable in circuit simulation}.

The paper establishes three main results. First, a thermal lag
$\varepsilon$ and an effective switching number $\Theta$
organize the model's response across variations in its physical
parameters. The latter combines the isothermal drift per
half-period, the ambient-temperature mobility correction,
and a closed-form approximation to the cycle-averaged Arrhenius
enhancement. Across $2.5\times10^{5}$ simulations spanning five
parameters, including thermal capacitance over ten decades,
$\Theta$ organizes the state-variable excursion in the tested
quasi-static domain with a robust relative scatter of $4\,\%$
and no fitted scaling constant
(\cref{sec:dimensionless}). Mapping the
$(\varepsilon,\Theta)$ plane at a fixed self-heating budget
quantifies reduction errors and bounds the region reachable
under the specified drive and geometry constraints.
The approximation for $\Theta$, together with the
sinusoidal-power proxy
$\varepsilon_{\rm est}=2\pi f\Rth\Cth$, provides an a priori
guide to these regimes; the trajectory-based thermal-lag
criterion determines the applicability of the proxy.

Second, the stroboscopic map of the quasi-statically reduced
model is proved to be strictly increasing on $[0,1]$,
structurally excluding period-doubling and chaos in that regime
\new{under current drive}.
\new{A unique attracting period-1 orbit is identified numerically
over the tested parameter ranges with the Biolek window.}
Its Floquet multiplier remains inside the unit circle and
determines the asymptotic settling time required for reliable
stroboscopic analysis (\cref{sec:poincare}).

Third, a variance-based sensitivity analysis with bootstrap
intervals over $(I_0,\Tamb,\Ea,\Rth)$ identifies thermal
resistance as a first-order contributor to self-heating.
For the logarithm of the loop area, used to improve the
conditioning of the variance estimates, activation energy
outranks thermal resistance (\cref{sec:sobol}).
Independent capacitance sweeps confirm the weak influence
of $\Cth$ in the quasi-static regime, consistent with its
absence from $\Theta$. \new{Within the sampled ranges, the effective thermal resistance
is therefore the more influential of the two thermal
parameters for the model's self-heating predictions. This is a parametric-sensitivity result
for the specified model, not a general experimental
prescription, although it indicates where calibration effort
is most informative.} The numerical predictions concern
the specified compact model with representative parameters;
device-specific use requires calibration of its electrical
and thermal parameters.

\Cref{sec:model} presents the model,
\cref{sec:dimensionless} develops the scaling and regime map,
and \cref{sec:numresults} reports the numerical verification.
\Cref{sec:qs-reduction} evaluates the quasi-static reduction,
\cref{sec:comparison} compares ATDM with existing compact
models, and \cref{sec:sobol} presents the sensitivity analysis.
\Cref{sec:limitations,sec:conclusion} discuss the limitations
and conclusions. Supporting verifications are collected in
\crefrange{app:equations}{app:veriloga}.

\section{The ATDM compact model}
\label{sec:model}

\subsection{Physical picture and modeling assumptions}

\begin{figure}[ht!]
\centering
\includegraphics[width=7cm,height=3cm]{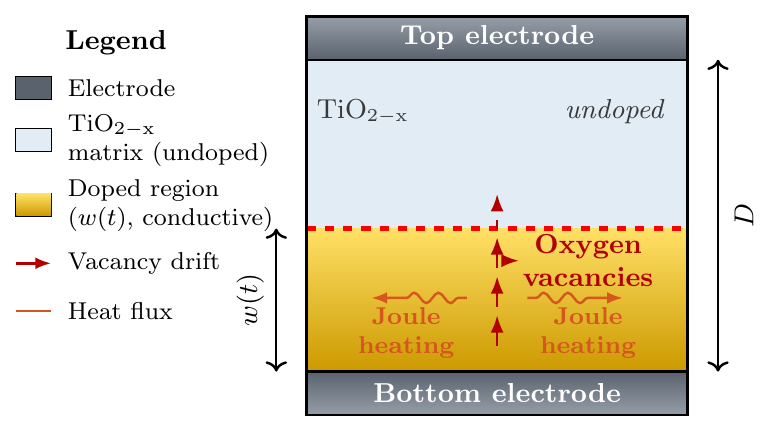}
\caption{Conceptual schematic motivating ATDM. The doped, more conductive
region (yellow, width $w(t)$) grows into the less-doped \ce{TiO2-x} region
(blue) in the effective moving-boundary picture. The sole electrical state is
$x(t)=w(t)/D$; the drawing is neither a measured nor a reconstructed filament
geometry. The lumped thermal node represents effective heat generation in, and
evacuation from, the electrically active region.}
\label{fig:device}
\end{figure}

The device is a metal--insulator--metal structure with a thin layer of
oxygen-deficient titanium dioxide (\ce{TiO2-x}) between two electrodes
(\cref{fig:device}).

Oxygen-vacancy redistribution is represented by the
one-dimensional width $w(t)$ of a vacancy-rich region: increasing $w$ lowers
the resistance in the Strukov series-resistor construction. This effective
moving-boundary picture \cite{Strukov2008,Pickett2009,Yang2013,Waser2007}
neither reconstructs a device microstructure nor resolves filament geometry
\cite{Li2012a}. The active region is a lumped electrothermal node with
effective $\Rth$, $\Cth$ \cite{Cahill2014}, whose baseline values are
order-of-magnitude estimates for a timescale, not measured device properties.

An increase in current produces additional Joule heating; the resulting
temperature rise modifies the ionic mobility, which changes the
moving-boundary dynamics and hence the resistance, which in turn sets the
Joule power---closing a nonlinear electrothermal loop. The model rests on the
following assumptions: (i) the active region is one spatially uniform
temperature $T(t)$, replacing the distributed heat equation by a single
thermal node ($\Rth$, $\Cth$); (ii) oxygen-vacancy migration dominates
switching and is thermally activated; (iii) electrical and thermal dynamics
are solved simultaneously; (iv) mechanical stress, stochastic fluctuations, and
structural degradation are neglected; (v) radiative and thermoelectric effects
are negligible against Joule heating; (vi) material parameters are constant
except where an explicit temperature dependence is introduced below.

\subsection{Governing equations}
\label{sec:tr3a}

Following Strukov \textit{et al.} \cite{Strukov2008}, the internal state
$x(t)=w(t)/D\in[0,1]$ is the normalized length of the vacancy-rich region,
$D$ the oxide thickness, and the total resistance is
\begin{equation}
R(x)=\Ron x+\Roff(1-x).
\label{eq:Rmem}
\end{equation}
The device is driven by $i(t)=I_0\sin(2\pi ft)$. As in most drift-based
compact models the linear drift equation is inaccurate near the physical
boundaries, so the Biolek window \cite{Biolek2009} is used,
\begin{equation}
F(x,i)=1-\left(x-H(i)\right)^{2p},\quad
H(i)=\begin{cases}0,& i>0\\ 1,& i\le0,\end{cases}
\label{eq:biolek}
\end{equation}
with $p$ controlling the sharpness of the boundary effect. The sign
convention is that of Strukov \textit{et al.}: a positive current increases
$x$, driving the device toward its low-resistance state (SET), and a negative
current decreases it (RESET). Accordingly $H(i)$ selects the boundary
\emph{toward which} the current drives $x$, so that $F\to0$ there and the
drift is progressively suppressed, while $F\to1$ at the opposite boundary,
from which the trajectory is moving away. The window is therefore not
symmetric in $x$: it switches branch with the sign of $i$, and
\cref{app:robustness} quantifies which of the results below depend on that
choice.

The classical model assumes a constant mobility $\muv=\mu_0$. Since vacancy
migration is thermally activated
\cite{Janotti2010,Morgan2010,DeSouza2015,DeLile2023}, ATDM instead uses an
Arrhenius law,
\begin{equation}
\muv(T)=\mu_0\exp\!\left[-\frac{\Ea}{\kb}\left(\frac1T-\frac1{T_0}\right)\right],
\label{eq:arrhenius}
\end{equation}
with $\mu_0$ the mobility at the reference temperature $T_0$ and $\Ea$ the
activation energy. This is the first of the two extensions. The second is that
temperature is an independent dynamical variable, governed by the first law
applied to the active region. The complete model reads
\begin{equation}
\left\{
\begin{aligned}
\frac{dx}{dt} &= \frac{\muv(T)\Ron}{D^2}\,i(t)\,F(x,i),\\[3pt]
\Cth\frac{dT}{dt} &= i^2R(x)-\frac{T-\Tamb}{\Rth},\\[3pt]
R(x) &= \Ron x+\Roff(1-x).
\end{aligned}
\right.
\label{eq:system3a}
\end{equation}
Whether the extra state must actually be integrated depends on the ratio of
thermal to excitation timescales; retaining the full two-state system lets
that question be answered rather than assumed
(\cref{sec:quasistatic-validation-3a}).

\subsection{Parameters and their status}
\label{sec:parameters}

Parameters are taken from experimentally grounded literature wherever
possible; those introduced by the electrothermal formulation are identified as
such. \Cref{tab:params} lists them\new{, and \cref{tab:nomenclature} collects
the symbols used throughout}. Only three quantities are new relative to
Strukov's model---the dynamic temperature $T(t)$, the Arrhenius mobility
$\muv(T)$, and the thermal network $(\Rth,\Cth)$---and only the last is a
modeling assumption in the strict sense.

All governing equations use absolute temperature in kelvin, as the Arrhenius
law and the thermal balance require. Ambient temperature is reported in
degrees Celsius wherever it appears as a swept parameter or axis label, since
that is the readable convention for a range centered on room temperature; the
reference is $T_0=293$~K throughout, and no equation is ever evaluated in
Celsius.

\begin{table*}[!tb]
\centering
\footnotesize
\caption{ATDM parameters. $\Rth$ and $\Cth$ are order-of-magnitude estimates
derived from bulk properties and an assumed active geometry, not values
measured on a device; $\Rth$ in particular is an \emph{effective} thermal
resistance omitting boundary resistances, electrode and substrate spreading,
and any localization of the current path (\cref{sec:thermal-interpretation}).}
\label{tab:params}
\begin{tabular}{llll}
\toprule
Parameter & Symbol & Value & Origin\\
\midrule
Oxide thickness & $D$ & $10$~nm & \cite{Strukov2008}\\
Active cross-sectional area & $A$ & $10\times10$~nm$^2$ & assumption\\
Low-resistance state & $\Ron$ & $100\,\Omega$ & \cite{Strukov2008}\\
High-resistance state & $\Roff$ & $16$~k$\Omega$ & \cite{Strukov2008}\\
Reference ionic mobility & $\mu_0$ & $10^{-14}$~m$^2$V$^{-1}$s$^{-1}$ & \cite{Strukov2008}\\
Reference temperature & $T_0$ & $293$~K & present work\\
Thermal conductivity & $\kappa$ & $1.6$~W\,m$^{-1}$K$^{-1}$ & \cite{Mun2007TiO2ThermalCond,Ok2018}\\
Mass density (rutile) & $\rho$ & $4250$~kg\,m$^{-3}$ & standard\\
Molar heat capacity & $C_{p,\rm m}$ & $55$~J\,mol$^{-1}$K$^{-1}$ & \cite{Smith2009}\\
Molar mass (\ce{TiO2}) & $M$ & $79.87$~g\,mol$^{-1}$ & standard\\
Specific heat capacity & $c_p=C_{p,\rm m}/M$ & $689$~J\,kg$^{-1}$K$^{-1}$ & derived\\
Effective thermal resistance & $\Rth$ & $6.25\times10^{7}$~K/W & $D/(\kappa A)$\\
Effective thermal capacitance & $\Cth$ & $2.93\times10^{-18}$~J/K & $\rho c_p AD$\\
Activation energy & $\Ea$ & $0.7$~eV & representative \cite{Morgan2010,Janotti2010}\\
\bottomrule
\end{tabular}
\end{table*}

The baseline thermal time constant, $\tau_{\rm th}=\Rth\Cth\approx
1.8\times10^{-10}$~s, is orders of magnitude smaller than any excitation
period considered here. That estimate supports the quasi-static reduction; it
does not establish the absolute temperature rise of a physical device. The
activation energy $\Ea=0.7$~eV lies within, but does not pin down, the
computed range $0.19$--$0.82$~eV for vacancy migration in rutile \ce{TiO2}
\cite{Morgan2010,Janotti2010,Morgan2009}.

\subsection{The effective thermal resistance}
\label{sec:thermal-interpretation}

The estimate $\Rth=D/(\kappa A)$ is a transparent baseline for
one-dimensional conduction across the assumed active volume, not a thermal
model of a metal--oxide--metal stack: boundary resistances, heat spreading in
electrodes and substrate, temperature-dependent material properties, and the
true electrically active area are all absorbed into one effective parameter.
Filamentary conduction is commonly reported over a cross-section far smaller
than the lithographic electrode area, and since $\Rth\propto1/A$ a smaller
active area would, in isolation, raise $\Rth$. The other unresolved factors
can act either way, so the baseline may under- or overestimate real
self-heating rather than being biased in one direction.

In the quasi-static regime derived below the dependence is explicit,

\begin{equation}
\left\{
\begin{aligned}
\Delta T_{\rm QS}(t)\equiv T_{\rm QS}(t)-\Tamb&=\Rth P_{\rm Joule}(t),
\\
\frac{\partial \Delta T_{\rm QS}}{\partial \Rth}&=P_{\rm Joule}(t),
\label{eq:Rth-sensitivity}
\end{aligned}
\right.
\end{equation}
so that at fixed instantaneous electrical state the relative sensitivity of
the temperature rise to $\Rth$ is exactly unity. This makes $\Rth$, not
$\Cth$, the principal uncalibrated thermal parameter of the low-frequency
regime.

\Cref{eq:Rth-sensitivity} holds $x(t)$ fixed, however, and therefore captures
only the direct effect. Integrating the full model at the baseline operating
point over $\Rth\in[0.1,10]\times$ its nominal value---a deliberately
hypothetical range, chosen to expose behavior rather than to assert bounds---
gives the comparison of \cref{tab:rth-sweep}. The deviation from the linear
extrapolation is systematic rather than noise: negative below the nominal
$\Rth$, positive above, saturating near $+21\,\%$ once
$\Rth\gtrsim2\times$ baseline. The mechanism is the excursion $\Delta x$
itself: weaker self-heating narrows the range explored by $x(t)$, so $R(x)$
spends less time near $\Roff$ and $P_{\rm Joule}$ falls below the
fixed-trajectory estimate---and conversely at high $\Rth$, until $\Delta x$
saturates against the window and the deviation plateaus. \Cref{eq:Rth-sensitivity}
is thus a useful local estimate, not a substitute for propagating $\Rth$
through the coupled model, which is what \cref{sec:sobol} does globally. The
numerical results below are accordingly conditional response maps around the
baseline $\Rth$; absolute temperatures require thermal calibration.

\begin{table}[!t]
\centering
\footnotesize
\caption{Numerical $\Delta T_{\max}$ against the linear extrapolation of
\cref{eq:Rth-sensitivity} (fixed nominal-$\Rth$ trajectory), swept over a
deliberately hypothetical range of $\Rth$ at $I_0=4\,\mu$A, $f=0.015$~Hz,
$\Ea=0.7$~eV.}
\label{tab:rth-sweep}
\begin{tabular}{ccccc}
\toprule
$\Rth/R_{\rm th,nom}$ & $\Delta x$ & numerical (K) & linear (K) & dev.\ (\%)\\
\midrule
$0.1$  & $0.739$ & $1.0$   & $1.3$   & $-24.5$\\
$0.5$  & $0.846$ & $5.6$   & $6.6$   & $-15.8$\\
$1.0$  & $0.935$ & $13.2$  & $13.2$  & $0.0$\\
$2.0$  & $0.976$ & $31.8$  & $26.4$  & $+20.5$\\
$5.0$  & $0.993$ & $79.9$  & $66.1$  & $+21.0$\\
$10.0$ & $0.997$ & $159.5$ & $132.2$ & $+20.6$\\
\bottomrule
\end{tabular}
\end{table}

\section{Dimensionless formulation and regime map}
\label{sec:dimensionless}

\Cref{eq:system3a} carries seven physical parameters ($\mu_0$, $\Ron$,
$\Roff$, $D$, $\Rth$, $\Cth$, $\Ea$) beyond the operating point
$(I_0,f,\Tamb)$. Reported in those variables, every result below would be a
statement about one parameter set. This section shows that the switching
response is organized by \emph{two} dimensionless groups, so that the results
become statements about the model itself.

\subsection{Three groups, no fitted constant}
\label{sec:groups}

\paragraph{Switching number.} Integrating the drift equation over a
half-period at constant mobility $\mu_0$ gives the excursion the isothermal
drift would produce,
\begin{equation}
\Pi \;\equiv\; \frac{\mu_0\Ron I_0}{\pi f D^{2}},
\label{eq:Pi}
\end{equation}
a pure number. $\Pi\ll1$ leaves $x$ frozen and the loop degenerate;
$\Pi\gg1$ drives $x$ into the boundaries of the window \cref{eq:biolek};
$\Pi\approx1$ is the nonsaturating regime. At the operating point selected
empirically in \cref{sec:validation-ATDM} ($I_0=4\,\mu$A, $f=0.015$~Hz) one
finds $\Pi=0.85$: the hand-tuned choice is recovered as an analytical
condition.

That $\Pi$ is the correct group in the isothermal limit can be checked
directly rather than argued. Writing $\tau=ft$ shows that at constant mobility
the drift depends on $I_0$ and $f$ only through their ratio, so every pair
$(I_0,f)$ sharing $I_0/f$ must produce the same orbit;
\cref{app:scaling} confirms this to a relative spread of $8\times10^{-6}$
across a factor of four in current and fourteen values of $I_0/f$, with a
measured slope $\Delta x=0.00298\,I_0/f$ ($I_0/f$ in $\mu$A/Hz) corresponding
to a mean window value $\langle F\rangle=0.936$. The same test shows why one
group cannot suffice: the current enters a second time through the Joule
power, which scales as $I_0^{2}$, so at fixed $I_0/f$ doubling $I_0$
quadruples the temperature rise and the single curve splits into three,
ordered by current.

\paragraph{Ambient offset.} The Arrhenius law is referenced to $T_0$, so a
sweep of $\Tamb$ rescales the mobility before any self-heating

\begin{equation}
G_{\rm amb}\equiv
\exp\!\left[
-\frac{\Ea}{\kb}
\left(
\frac{1}{\Tamb}-\frac{1}{T_0}
\right)
\right].
\label{eq:Gamb}
\end{equation}

\paragraph{Feedback strength and its cycle average.} Evaluating the Arrhenius
exponent on the quasi-static rise $\Delta T_{\rm QS}=\Rth I_0^{2}\bar R$, with
$\bar R=(\Ron+\Roff)/2$, defines the feedback strength

{\footnotesize
\begin{equation}
\Lambda\;\equiv\;\frac{\Ea}{\kb}
\left(\frac{1}{\Tamb}-\frac{1}{\Tamb+\Delta T_{\rm QS}}\right)
\;\xrightarrow[\Delta T_{\rm QS}\ll\Tamb]{}\;
\frac{\Ea\,\Delta T_{\rm QS}}{\kb\Tamb^{2}}.
\label{eq:Lambda}
\end{equation}
}
The exact form is retained because $\Delta T_{\rm QS}$ reaches a sizeable
fraction of $\Tamb$ in \cref{sec:regimemap}, where the linearization
overestimates the enhancement by orders of magnitude. At the baseline
$\Lambda=0.74$, hence a peak enhancement $e^{\Lambda}=2.1$.

What enters the drift over a full cycle is not the peak enhancement but its
mean. In the quasi-static regime $\Delta T(t)\simeq\Delta T_{\rm QS}
\sin^{2}(2\pi ft)$, so the cycle-averaged Arrhenius factor is available in
closed form,
\begin{equation}
\langle\Gamma\rangle\;\equiv\;
\big\langle e^{\Lambda\sin^{2}\theta}\big\rangle_\theta
\;=\;e^{\Lambda/2}\,\mathcal{I}_0\!\left(\tfrac{\Lambda}{2}\right),
\label{eq:Gamma}
\end{equation}
with $\mathcal{I}_0$ the modified Bessel function of the first kind of order
zero---written in calligraphic type to avoid any confusion with the current
amplitude $I_0$.
No constant is fitted anywhere in \crefrange{eq:Pi}{eq:Gamma}.

\Cref{eq:Gamma} is a controlled approximation, not an identity. \new{It
rests on two distinct simplifications. First, the instantaneous Arrhenius
exponent, $(\Ea/\kb)\,[1/\Tamb-1/(\Tamb+\Delta T(t))]$, is replaced by
$\Lambda\,\Delta T(t)/\Delta T_{\rm QS}$; the two coincide when
$\Delta T=0$ and $\Delta T=\Delta T_{\rm QS}$, but the exact exponent is larger
by the factor $(\Tamb+\Delta T_{\rm QS})/(\Tamb+\Delta T)\ge1$ in between,
so this step underestimates the enhancement as $\Delta T_{\rm QS}/\Tamb$
grows. Second,} it assumes
$\Delta T(t)=\Delta T_{\rm QS}\sin^{2}(2\pi ft)$, whereas the power in the
full model is $i^2R(x(t))$ and therefore carries the state's own waveform.
\Cref{tab:gamma-check} in \cref{app:robustness} compares the closed form with
the exact cycle average of $\muv(T(t))/\mu_0$ on settled orbits. The agreement
is $1.2\,\%$ in median for $\Lambda\lesssim1$, which covers the baseline and
most of the grid, and degrades systematically---the closed form
\emph{under}-estimating the enhancement---as the feedback strengthens:
$-16\,\%$ at $\Lambda=1.14$ and $-52\,\%$ at $\Lambda=2.75$. Because $\Theta$
serves as an ordering variable rather than as a predicted value, such an error
displaces a point along the master curve rather than away from it, and the
residual scatter quoted below is measured on the ensemble as it stands,
strongly driven points included. Read quantitatively, however, $\Theta$ is a
weak-to-moderate-feedback construction.

\paragraph{Collapse variable.} The three combine into a single effective
switching number,
\begin{equation}
\boxed{\;\Theta\;\equiv\;\Pi\,G_{\rm amb}\,\langle\Gamma\rangle\;}
\label{eq:Theta}
\end{equation}
which, with the thermal lag $\varepsilon=\tau_{\rm th}/\tau_P$ of
\cref{eq:epsilon-3a}, spans the model's behavior.

\subsection{Data collapse}
\label{sec:collapse}

\begin{figure*}[ht!]
\centering
\includegraphics[width=16cm,height=4.5cm, trim = 0 0 0 0, clip]{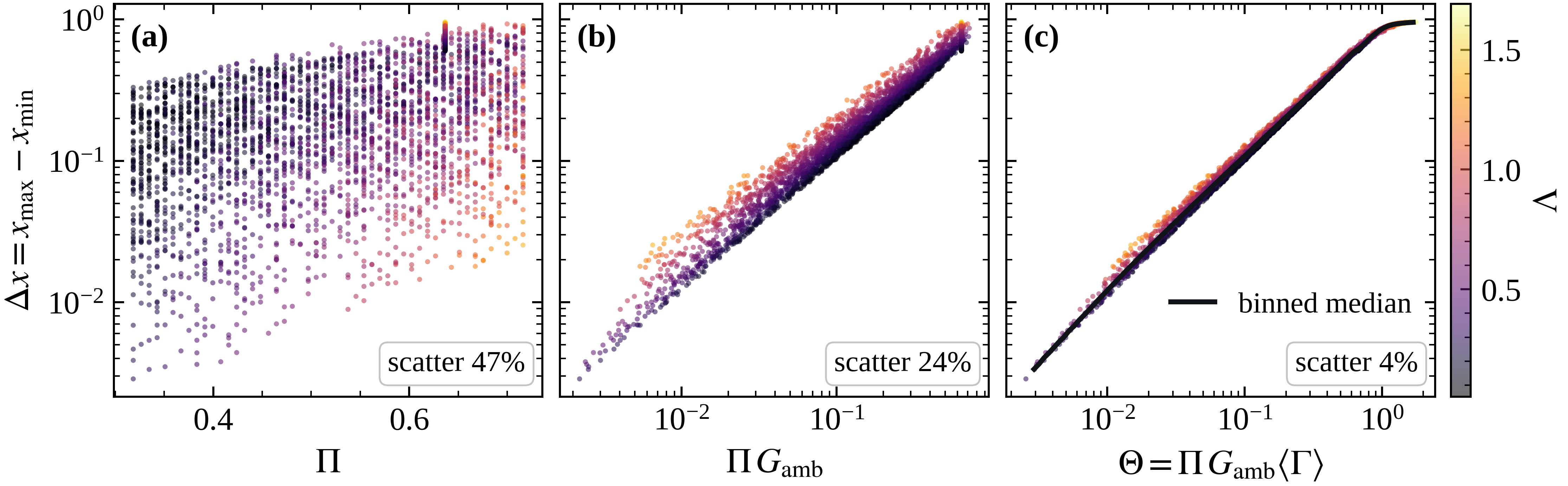}
\caption{Collapse of the state-variable excursion onto a single dimensionless
group, over $2.5\times10^{5}$ stored simulations from two independent
full-factorial grids. (a) Against the isothermal switching number $\Pi$ alone.
(b) Against $\Pi G_{\rm amb}$, adding the ambient Arrhenius offset.
(c) Against $\Theta=\Pi G_{\rm amb}\langle\Gamma\rangle$, adding the
cycle-averaged self-heating enhancement of \cref{eq:Gamma}; the black curve is
the binned median. Color encodes the feedback strength $\Lambda$. The
annotated residual scatter falls from $47\,\%$ to $4\,\%$ with no fitted
constant.}
\label{fig:collapse}
\end{figure*}

\Cref{fig:collapse} tests \cref{eq:Theta} on two full-factorial grids---
$2.5\times10^{5}$ simulations in total, spanning $I_0$, $\Tamb$, $\Ea$, $\Rth$,
and $\Cth$, the last over ten decades. No new integration is performed: the
stored outputs of \cref{sec:sobol-maps,sec:Cth-3a} are replotted against
$\Theta$.

The residual scatter is defined as follows: the abscissa is cut into $40$
logarithmically spaced bins, the median of $\Delta x$ is taken in each, and
the figure quoted is the median over bins of the interquartile range divided
by twice the bin median---a robust relative dispersion, insensitive to the
tails. Saturated points, where the window pins $\Delta x$ at unity and any
candidate variable would agree trivially, are excluded. On that measure the
excursion scatters by $47\,\%$ against $\Pi$ alone; multiplying by
$G_{\rm amb}$ brings it to $24\,\%$ and including $\langle\Gamma\rangle$ to
$4\,\%$. The defensible statement is therefore that $\Theta$ gives an \new{empirical}
one-parameter organization of the settled excursion over the quasi-static
domain tested here, and not a universal law: the black curve of
\cref{fig:collapse}(c) is a binned median rather than an analytically derived
function, and what it organizes is $\Delta x$ alone. Two consequences
deserve stating. First, $\Cth$ does not appear in $\Theta$, yet the second
grid sweeps it over ten decades: \new{the weak influence of $\Cth$ in the
tested quasi-static regime is consistent with its absence from $\Theta$ and
is confirmed by the independent capacitance sweeps of \cref{sec:Cth-3a}.} Second, the collapse does not extend to
the loop area, whose residual scatter remains near $60\,\%$ on a logarithmic
scale---consistent with the interaction-dominated character that
\cref{sec:sobol} measures, and identifying that output as requiring a second
group the present scaling does not supply.

\subsection{Regime map}
\label{sec:regimemap}

Spanning the $(\varepsilon,\Theta)$ plane requires care. Sweeping the two
obvious physical knobs, $f$ and the active area $A$, does \emph{not} fill it:
$\varepsilon\propto f$ while $\Pi\propto1/f$, so the whole grid projects onto
the single line $\Theta\propto1/\varepsilon$. The plane is therefore
parametrized directly. The peak quasi-static temperature rise is fixed at a
budget $\Delta T_{\rm b}=150$~K---large enough for the feedback to act, small
enough for a constant-property lumped description to remain defensible---and
each target $(f,\Theta)$ then determines the drive and geometry that realize
it,
\begin{equation}
\Pi=\frac{\Theta}{G_{\rm amb}\langle\Gamma\rangle},\;
I_0=\frac{\Pi\,\pi f D^{2}}{\mu_0\Ron},\;
\Rth=\frac{\Delta T_{\rm b}}{I_0^{2}\bar R},\;
A=\frac{D}{\kappa\Rth},
\label{eq:realise}
\end{equation}
with $\Cth=\rho c_p A D$. The last two are not independent: $\Rth$ and $\Cth$
together define the geometry, and $\tau_{\rm th}=\rho c_p D^{2}/\kappa$ is
area-independent precisely because both scale with $A$. Varying one without
the other would silently change $\tau_{\rm th}$---invisible while
$\varepsilon\ll1$, an order of magnitude on $\Delta x$ once
$\varepsilon\sim1$.

\begin{figure*}[!ht]
\centering
\includegraphics[width=16.0cm,height=4cm, trim = 0 0 0 0, clip]{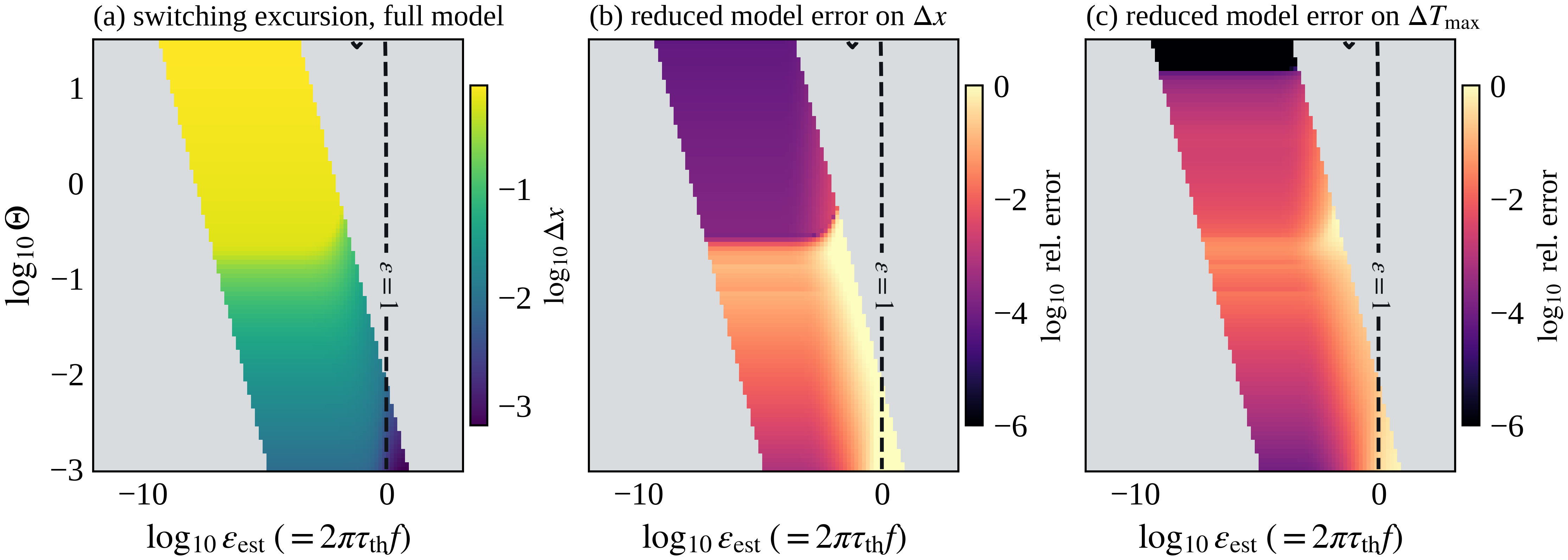}
\caption{Regime map in the $(\varepsilon,\Theta)$ plane at a fixed peak
self-heating budget of $150$~K, each target realized by the drive and geometry
of \cref{eq:realise}. Gray marks the physically unreachable part of the plane.
(a) Switching excursion of the full model. (b) Relative error of the
quasi-static reduced model on $\Delta x$. (c) The same on $\Delta T_{\max}$.
The dashed line is the measured $\varepsilon=1$ boundary; the abscissa is its
analytic proxy $2\pi\tau_{\rm th}f$: for $P=P_{\max}\sin^{2}(2\pi ft)$ one has
$|\dot P|_{\max}=2\pi fP_{\max}$ and hence $\tau_P=1/(2\pi f)$, so the
abscissa is an exact rescaling of the swept frequency. The same proxy places
$\varepsilon=1$ at $870$~MHz for the nominal geometry, consistent with
\new{\cref{sec:qs-frequency-range}}.}
\label{fig:regimemap}
\end{figure*}

Points whose implied geometry falls outside
$A\in[1\,\mathrm{nm}^{2},1\,\mathrm{mm}^{2}]$ or
$I_0\in[1\,\mathrm{nA},1\,\mathrm{A}]$ are grayed in \cref{fig:regimemap}.
That boundary is a result in itself: at a $150$~K budget only $38\,\%$ of the
plane is physically reachable, and the accessible band is what limits how far
into the thermally-lagged regime a real device can be driven. Within it, the
reduced one-state formulation of \cref{sec:qs-reduction} reproduces the full
model to a median $0.07\,\%$ on $\Delta x$ for $\varepsilon<10^{-2}$ and
departs from it by a median $560\,\%$ for $\varepsilon>1$: the threshold
located in \cref{sec:qs-frequency-range} is not a property of one frequency
sweep but a line of the plane, with the error quantified on either side.
Lowering the budget to $50$~K removes the intersection of the accessible band
with $\varepsilon=1$ altogether---at that thermal budget the quasi-static
reduction is never violated anywhere a device can physically operate.

\begin{figure*}[!tb]
\centering
\includegraphics[width=0.96\textwidth]{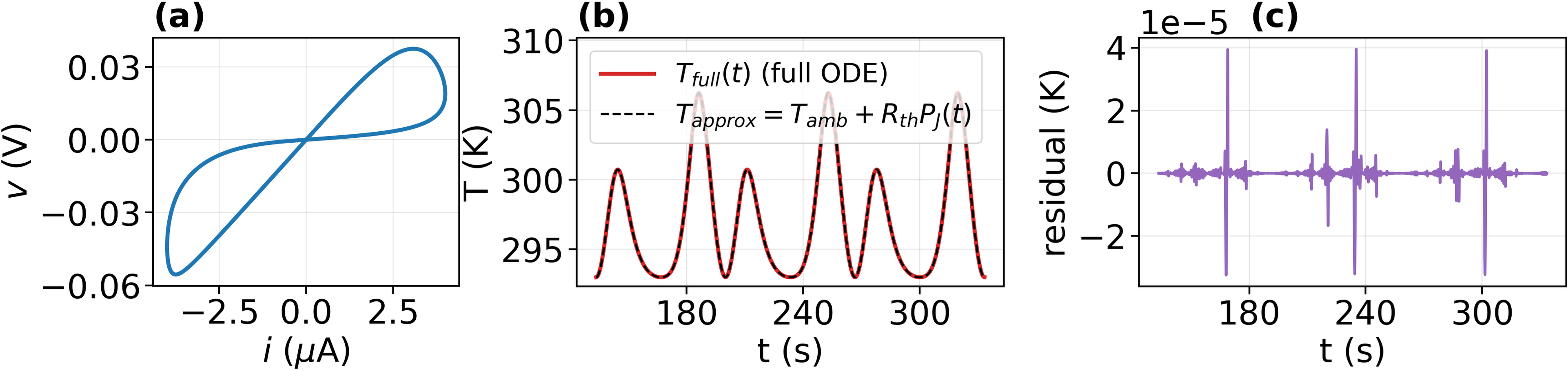}
\caption{(a) Current--voltage hysteresis loop of ATDM at the baseline
operating point ($I_0=4\,\mu$A, $f=0.015$~Hz, $\Ea=0.7$~eV), over the last
simulated period. (b) Full thermal solution $T_{\rm full}(t)$ (red) against
the quasi-static approximation $\Tamb+\Rth P_{\rm Joule}(t)$ (black dashed)
over the last three forcing periods; the two overlap to within plotting
resolution. (c) Their difference on the same axis, below $40\,\mu$K
throughout, that is $3\times10^{-4}\,\%$ of the $13.2$~K peak temperature
rise---consistent with solver tolerance rather than with a breakdown of the
approximation.}
\label{fig:results3a}
\end{figure*}

\section{Numerical verification}
\label{sec:numresults}

\subsection{Mathematical and numerical properties}
\label{sec:tau}

ATDM is a nonlinear, periodically forced system with state
$\mathbf X=[x,T]^{\!\top}$. On each open current-sign interval $R(x)$ and
$\muv(T)$ are continuously differentiable for $T>0$ and the right-hand side is
locally Lipschitz. The Biolek window changes branch at $i=0$ and is therefore
not continuous in $i$, but the drift term $iF(x,i)$ has the same zero limit
from both sides, so the system is interpreted piecewise between current zero
crossings and continued uniquely across them. For $0\le x(0)\le1$ and
$T(0)>0$, the window makes $[0,1]$ forward invariant while the thermal
equation points toward positive temperatures at the lower boundary; both
bounds are monitored numerically to separate physics from solver overshoot.

The thermal time constant $\tau_{\rm th}=\Rth\Cth\approx1.8\times10^{-10}$~s
is many orders of magnitude below the excitation periods considered
($0.015$--$1$~Hz). Two consequences follow: the coupled system is stiff, and
is integrated throughout this work with the implicit Radau IIA method
\cite{Hairer1996II,Hairer2015} at $\mathrm{rtol}=10^{-6}$,
$\mathrm{atol}=10^{-9}$; and temperature tracks the instantaneous dissipated
power rather than exhibiting slow thermal hysteresis of its own.

\subsection{Baseline response and quasi-static reduction}
\label{sec:quasistatic-validation-3a}

ATDM is simulated under $i(t)=I_0\sin(2\pi ft)$ from thermal equilibrium.
Unless noted the operating point is $I_0=4\,\mu$A, $f=0.015$~Hz, chosen so
that $x(t)$ traverses most of $[0,1]$ within a half-cycle without saturating.
\Cref{fig:results3a}(a) shows the resulting $i$--$v$ trajectory: over the
settled period $x$ stays
within $[0.007,0.943]$ and the temperature rises by $\approx13$~K, giving a
visibly open pinched loop. Every temperature quoted in this work is a
model-level estimate for the nominal parameter set of \cref{tab:params}, and
in particular for the assumed $100$~nm$^2$ active area that fixes $\Rth$
through $D/\kappa A$; it is not a device-independent prediction, and a more
localized current path would raise it (\cref{sec:thermal-interpretation}).

In the limit $\tau_{\rm th}f\ll1$ the thermal balance reduces to the algebraic
relation
\begin{equation}
T(t)\;\simeq\;\Tamb+\Rth P_{\rm Joule}(t),
\label{eq:quasistatic}
\end{equation}
obtained by setting $\dot T\approx0$. The relevant small parameter is not
$\tau_{\rm th}f$ but
\begin{equation}
\varepsilon\;\equiv\;\frac{\tau_{\rm th}}{\tau_P},\qquad
\tau_P\;\equiv\;\frac{P_{\max}}{|\dot P_{\rm Joule}|_{\max}},
\label{eq:epsilon-3a}
\end{equation}
where $P_{\max}$ and $|\dot P_{\rm Joule}|_{\max}$ are both taken over the
\emph{last} forcing period of a settled orbit, so that $\tau_P$ measures the
established regime and not the initial transient; the derivative is evaluated
on the solver's output grid and its extremum taken in absolute value, so that
sign changes of $\dot P$ play no role. Defined this way $\tau_P$ is a property
of the trajectory and not of the integrator: at the baseline it is $8.8720$~s
at $\mathrm{rtol}=10^{-6}$ against $8.8709$~s at $10^{-9}$, four digits apart,
giving $\varepsilon\approx2.06\times10^{-11}$.

The reason for not writing $\varepsilon=\tau_{\rm th}f$ is that what the
thermal node must follow is the power, not the drive, and the two share a
timescale only for a sinusoid. \Cref{tab:tauP-check} in
\cref{app:robustness} repeats the measurement at fixed $f=0.015$~Hz for a
triangular and for a smoothed square current: $\tau_Pf$ falls from $0.133$ to
$0.058$ and then $0.011$, so $\varepsilon$ grows by an order of magnitude at
unchanged frequency. A criterion written on $f$ alone would call the three
equally quasi-static; the criterion written on $\tau_P$ separates them, which
is what the reduction requires.

\Cref{fig:results3a}(b,c) verifies this directly, comparing the full solution
against \cref{eq:quasistatic} computed \emph{a posteriori} from the same
$P_{\rm Joule}(t)$: the residual stays below $40\,\mu$K, that is
$3\times10^{-4}\,\%$ of the $13.2$~K peak rise---solver tolerance rather than
a physical departure. In the
baseline regime the algebraic relation is therefore the practically relevant
thermal formulation, and the full dynamic equation is retained to define the
model outside it.

\subsection{Where the quasi-static reduction fails}
\label{sec:qs-frequency-range}

\begin{figure}[ht!]
\centering
\includegraphics[width=8.5cm,height=4cm, trim = 0 0 0 0, clip]{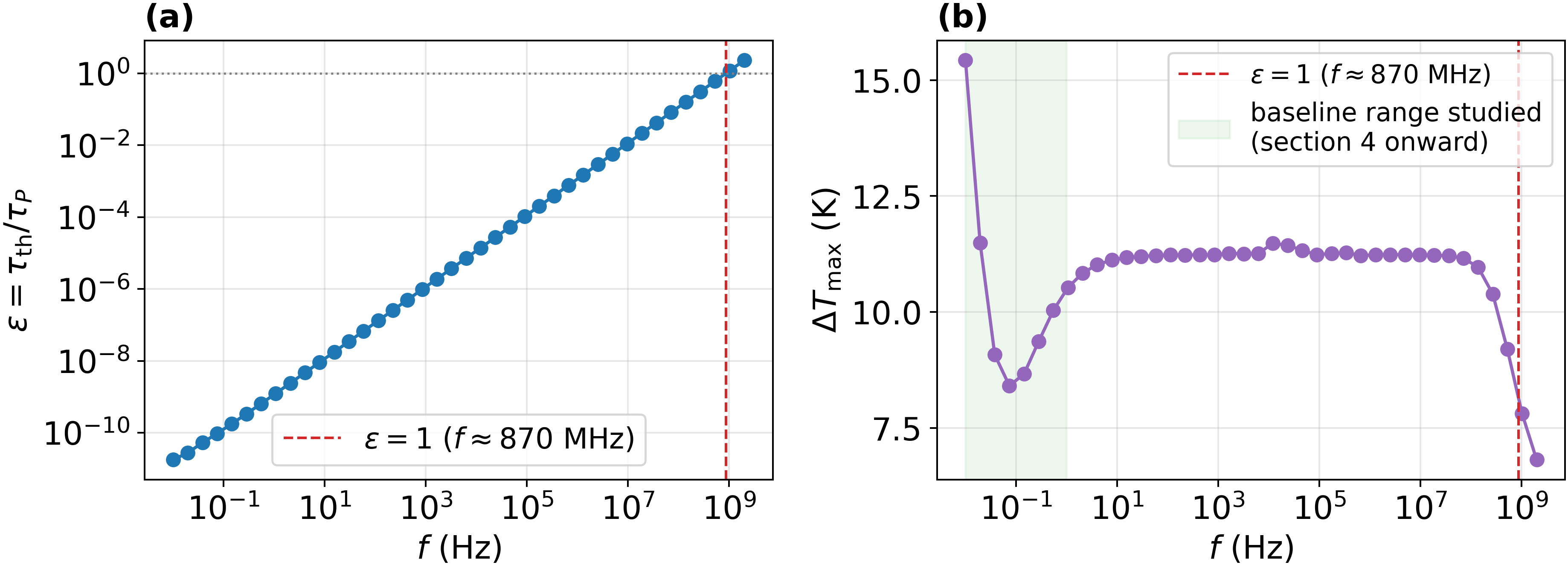}
\caption{Quasi-static parameter $\varepsilon=\tau_{\rm th}/\tau_P$ (a) and
peak self-heating $\Delta T_{\max}$ (b) against excitation frequency
($I_0=4\,\mu$A, $\Ea=0.7$~eV, $\Rth$ and $\Cth$ at baseline), swept over
eleven decades. $\varepsilon$ crosses unity at $f\approx870$~MHz;
$\Delta T_{\max}$ is flat across the quasi-static plateau and falls sharply
beyond that threshold. The green band marks the range studied elsewhere in
this work.}
\label{fig:qs-breakdown}
\end{figure}

The baseline lies deep inside the quasi-static regime by construction. At what
frequency does $\varepsilon$ actually approach unity, and how does that
frequency compare with real device operation? Sweeping $f$ over eleven decades
at $I_0=4\,\mu$A shows $\varepsilon$ growing linearly with $f$ and crossing
unity at $f\approx870$~MHz for the nominal geometry---a period of $1.15$~ns,
and a threshold that moves with $\tau_{\rm th}=\rho c_pD^2/\kappa$ rather than
being a property of \ce{TiO2} as such
(\cref{fig:qs-breakdown}(a)). The peak self-heating tracks the transition
(\cref{fig:qs-breakdown}(b)): after an initial variation at very low frequency
---a state-variable boundary effect, unrelated to thermal dynamics---
$\Delta T_{\max}$ plateaus at $\approx11.2$~K across seven decades before
falling to $7.8$~K by $1$~GHz, the thermal node losing the ability to track
the instantaneous Joule power.

\new{This breakdown frequency is a property of the specified parameter set,
not a prediction for a particular device. For context only, subnanosecond
SET/RESET switching ($105$--$120$~ps pulses) has been measured in a tantalum
oxide memristor \cite{Torrezan2011}, a different material and device
architecture, and Strukov and Williams identify conditions for
nanosecond-scale switching in exponential ionic-drift models
\cite{StrukovWilliams2009}. Such fast-switching regimes motivate locating
where the quasi-static reduction ceases to apply; they do not provide
experimental support for the thresholds computed here for \ce{TiO2}.}

\Cref{fig:qs-breakdown} locates the boundary through $\varepsilon$ and
$\Delta T_{\max}$; the regime map of \cref{sec:regimemap} quantifies the error
incurred on either side of it across the whole plane. Two numerical stress tests at single operating points, run far outside the
nominal parameter set and reported as asymptotic probes rather than as device
predictions, are given in \cref{app:beyond-qs}: at $1$ and $3$~GHz the reduced model overshoots the
temperature swing by factors of $2.5$ and $7$, and once the active area is
enlarged so that appreciable switching occurs at bounded temperature, the two
formulations diverge in $x(t)$ itself---different operating points and an
excursion overestimated by $2.6$--$3.2\times$---not only in temperature.

\subsection{Electrothermal feedback and frequency doubling}
\label{sec:joule-3a}

\begin{figure*}[!tb]
\centering
\includegraphics[width=17cm,height=4cm, trim = 0 0 0 0, clip]{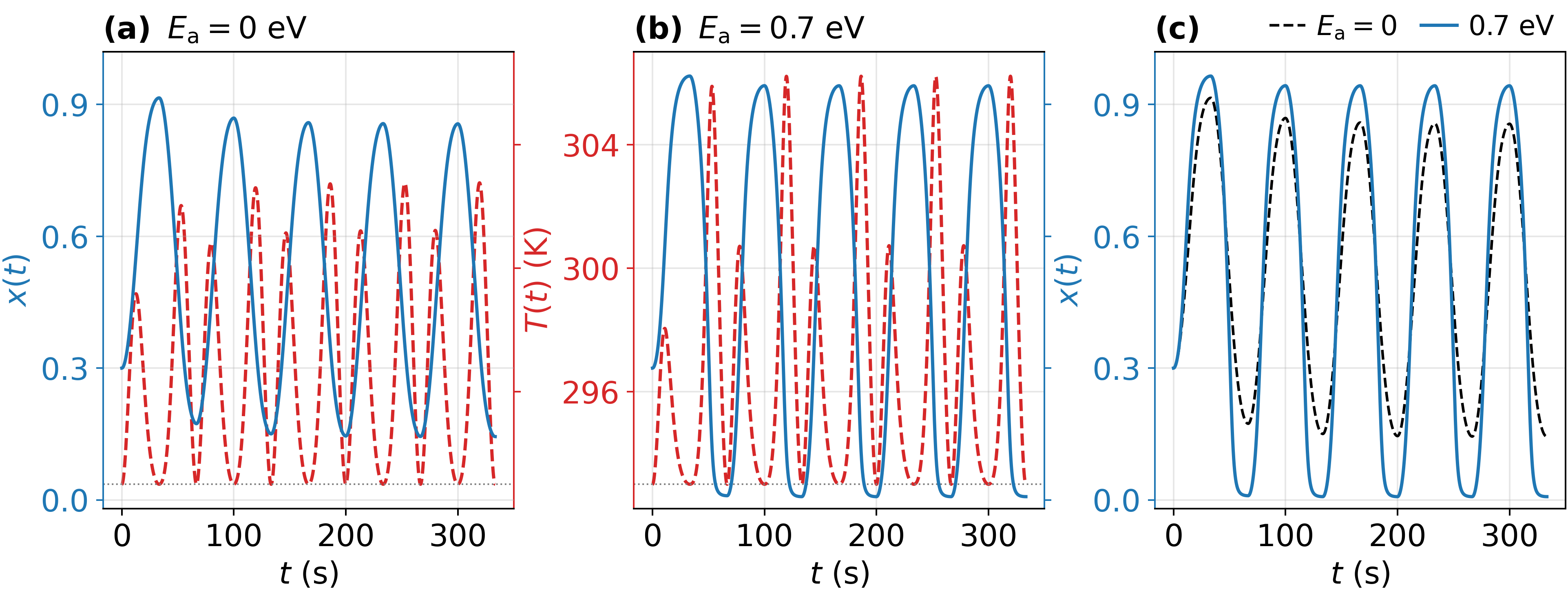}
\caption{Simultaneous evolution of the state variable $x(t)$ (blue) and
temperature $T(t)$ (red dashed) at $I_0=4\,\mu$A, $f=0.015$~Hz. Each vertical
axis carries the color of the quantity it measures, and an axis standing
between two panels serves both, so the panels are directly comparable.
(a) Isothermal limit, $\Ea=0$: $T(t)$ evolves but does not feed back on
$\muv$. (b) Thermally activated case, $\Ea=0.7$~eV. (c) The two state
trajectories superposed: the feedback widens the excursion from
$\Delta x=0.712$ to $0.935$ while the peak temperature rises from $302.8$ to
$306.2$~K.}
\label{fig:xT}
\end{figure*}

\Cref{fig:xT} shows $x(t)$ and $T(t)$ at the baseline operating point, with
and without thermal activation. \new{The temperature exhibits two unequal heating peaks per forcing
period, one in each current half-cycle. The Joule power is
$P_{\rm Joule}(t)=i^2(t)R(x(t))$, where
$i^2(t)=I_0^2[1-\cos(4\pi ft)]/2$ contains a component at $2f$.
The evolving resistance modulates this factor, so the heating
peaks need not coincide with the extrema of the current.
Their unequal heights reflect the different resistance
trajectories during the two half-cycles. According to
$R(x)=\Ron x+\Roff(1-x)$, $x\approx0$ corresponds to
$R\approx\Roff$, whereas $x\approx1$ corresponds to
$R\approx\Ron$; at a given current magnitude, the more
resistive state dissipates more power. In the quasi-static
regime considered here, the temperature closely follows
this modulated power, as verified in \cref{fig:results3a}(b,c).} The same doubling appears in the instantaneous Joule
power, which alternates \new{over the settled forcing periods} between \new{$\sim120$} and $\sim211$~nW peaks
(\new{\cref{fig:joule-T}}).

Comparing the two panels isolates the feedback itself. At $\Ea=0$ the
temperature still evolves---Joule heat is dissipated regardless---but does
not act on $\muv$, and the excursion is narrower: over the settled period
$x\in[0.144,0.856]$ against
$x\in[0.007,0.943]$, with peak temperatures $302.8$ and $306.2$~K. The feedback
widens $\Delta x$ from $0.712$ to $0.935$, a factor $1.31$, by letting $x(t)$
travel farther within the same half-cycle, which changes the time-averaged
$R(x)$ and hence the Joule power that drove it.

\subsection{Joint dependence on drive and ambient temperature}
\label{sec:joule-2D-3a}

\begin{figure}[ht!]
\centering
\includegraphics[width=9cm,height=4cm, trim = 0 0 0 2cm, clip]{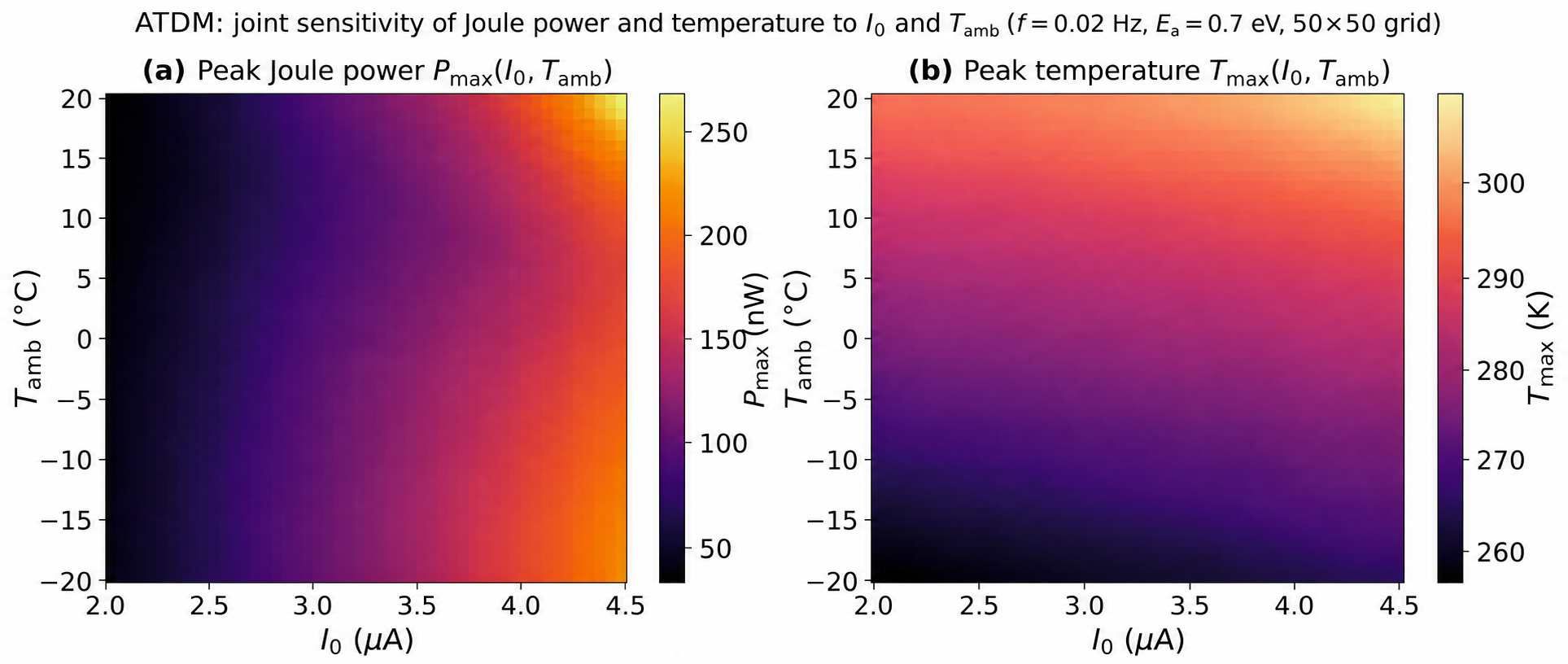}
\caption{Peak Joule power $P_{\max}(I_0,\Tamb)$ (a) and peak temperature
$T_{\max}(I_0,\Tamb)$ (b) over a $50\times50$ grid; $f=0.02$~Hz,
$\Ea=0.7$~eV, $T_0=293$~K. $P_{\max}$ is nonmonotonic in $\Tamb$ at fixed
$I_0$ (see text and \cref{tab:phase-lag}), while $T_{\max}$ increases
monotonically throughout.}
\label{fig:joule2D}
\end{figure}

\Cref{fig:joule2D} sweeps $I_0\in[2,4.5]\,\mu$A and
$\Tamb\in[-20,20]\,^\circ$C at $f=0.02$~Hz. The peak Joule power is
\emph{nonmonotonic} in $\Tamb$ at fixed $I_0$: it decreases from
$-20\,^\circ$C to a minimum near $6$--$12\,^\circ$C, then rises again---
verified smooth by a finer $21$-point sweep at $I_0=4\,\mu$A, ruling out a
numerical artifact.

\new{The resistance modulation behind this trend can be illustrated by the value
of $x$ at the instant the current peaks, $x_{i_{\max}}$, which sets
$P(t_{i_{\max}})=i_{\max}^{2}R(x_{i_{\max}})$; this power at the current
maximum is not identical to $P_{\max}$.}
\Cref{tab:phase-lag} shows $x_{i_{\max}}$ itself to be nonmonotonic, growing
from $0.330$ at $-20\,^\circ$C to $0.488$ near $12\,^\circ$C before falling to
$0.417$ at $20\,^\circ$C---even though the excursion amplitude increases
monotonically with $\Tamb$ throughout (\cref{sec:hysteresis-vs-Tamb-3a}).
Since $\Roff\gg\Ron$, $R(x_{i_{\max}})$ mirrors it inversely, producing the
U-shape. Physically this is a phase-lag effect: as $\muv(\Tamb)$ grows, $x(t)$
responds increasingly quickly to the forcing and the lag between the peaks of
$i(t)$ and $x(t)$ shrinks, so the value $x$ takes when $i$ peaks depends on an
evolving phase relationship rather than on amplitude alone. \new{In the quasi-static regime considered here,
$T_{\max}\simeq\Tamb+\Rth P_{\max}$, where
$P_{\max}=\max_t P_{\rm Joule}(t)$ over a settled forcing period.
Although $P_{\max}$ varies nonmonotonically with $\Tamb$,
the additive increase in ambient temperature outweighs the
decrease in self-heating where it occurs, so $T_{\max}$ rises
monotonically from approximately $266$ to $310$~K over the
reported sweep.}

\begin{table}[ht!]
\centering
\footnotesize
\caption{State variable and resistance at the instant of peak current,
$I_0=4\,\mu$A, $f=0.02$~Hz, $\Ea=0.7$~eV.}
\label{tab:phase-lag}
\begin{tabular}{lccc}
\toprule
$\Tamb$ & $x_{i_{\max}}$ & $R(x_{i_{\max}})$ & $P(t_{i_{\max}})$\\
\midrule
$253$~K ($-20\,^\circ$C) & $0.330$ & $10761\,\Omega$ & $172.2$~nW\\
$265$~K ($-8\,^\circ$C)  & $0.389$ & $9812\,\Omega$  & $157.0$~nW\\
$279$~K ($6\,^\circ$C)   & $0.484$ & $8309\,\Omega$  & $133.0$~nW\\
$285$~K ($12\,^\circ$C)  & $0.488$ & $8240\,\Omega$  & $131.8$~nW\\
$293$~K ($20\,^\circ$C)  & $0.417$ & $9366\,\Omega$  & $149.9$~nW\\
\bottomrule
\end{tabular}
\end{table}

\new{\Cref{tab:phase-lag} characterizes the power at the current maximum,
rather than the maximum power over the cycle; it illustrates
the resistance modulation underlying the trend in \cref{fig:joule2D}.}

\subsection{Activation energy and thermal capacitance}
\label{sec:Cth-3a}

\begin{figure}[ht!]
\centering
\includegraphics[width=8.7cm,height=4cm, trim = 0 0 0 0, clip]{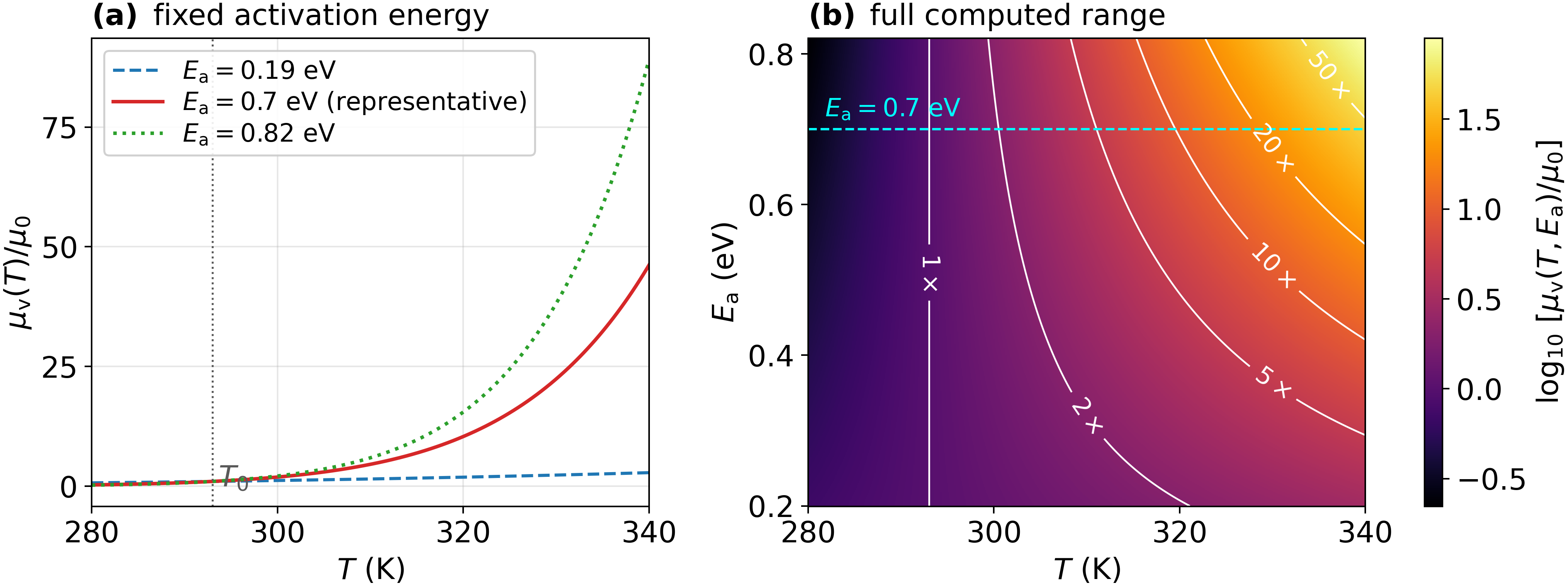}
\caption{Temperature dependence of the ionic mobility predicted by
\cref{eq:arrhenius}, normalized to its value $\mu_0$ at $T_0=293$~K.
(a) Three curves at the representative activation energy $\Ea=0.7$~eV (solid)
and the bounds of the computed range, $0.19$~eV (dashed) and $0.82$~eV
(dotted); the vertical dotted line marks $T_0$. (b) Continuous map of
$\muv(T,\Ea)/\mu_0$ on a logarithmic color scale over the same range, with
white contours at selected multiplicative factors and the cyan dashed line at
$\Ea=0.7$~eV.}
\label{fig:mobility}
\end{figure}

The temperature excursions above translate into a substantial modulation of
the ionic mobility. \Cref{fig:mobility}(a) plots $\muv(T)/\mu_0$ over the
range spanned by the simulations at $\Ea=0.7$~eV and at the two bounds of the
computed range, $0.19$ and $0.82$~eV. At $0.7$~eV the mobility reaches
$3.2\times\mu_0$ at $306$~K---the peak temperature of \cref{fig:xT}---and
$10.4\times\mu_0$ at $320$~K; at $0.19$~eV the same temperatures give only
$1.4\times$ and $1.9\times$. Panel (b) extends this to the full $(T,\Ea)$
plane: at $T=320$~K the enhancement varies from $\times2$ to $\times15$ across
the computed range of $\Ea$, a spread of the same order as the $\times1$ to
$\times10$ obtained by sweeping $T$ alone at fixed $\Ea$. The sensitivity to
where within that range the true activation energy lies is therefore
comparable to the sensitivity to temperature itself.

\begin{figure}[ht!]
\centering
\includegraphics[width=8.5cm,height=6cm, trim = 0 0 0 0, clip]{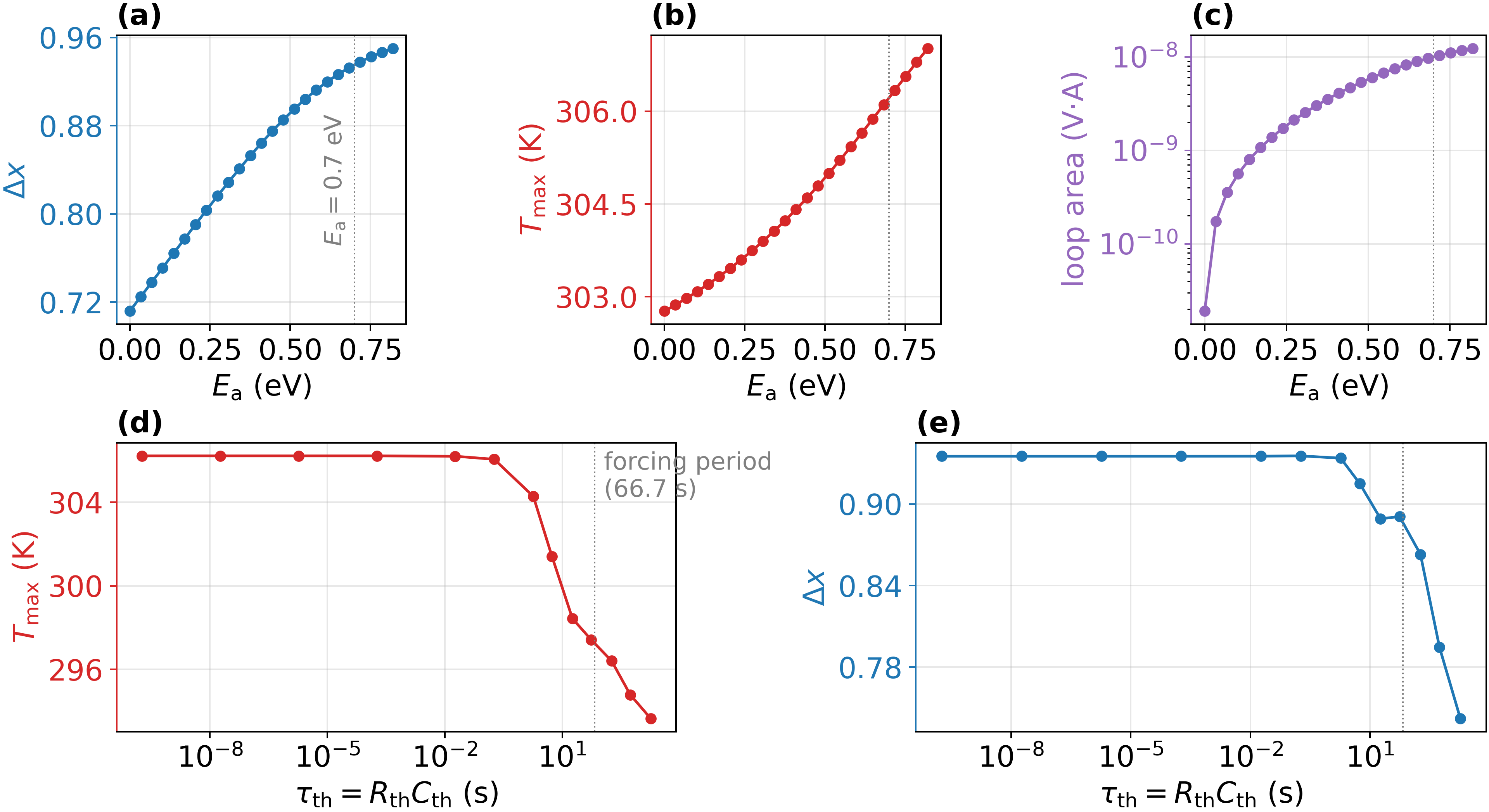}
\caption{(a--c) State-variable excursion $\Delta x$, peak temperature
$T_{\max}$, and hysteresis loop area---the last on a logarithmic ordinate,
since it varies by a factor $647$---as continuous functions of
$\Ea\in[0,0.82]$~eV at $I_0=4\,\mu$A, $f=0.015$~Hz; the dotted line marks the
representative $\Ea=0.7$~eV. (d, e) Peak temperature and excursion against the
thermal time constant $\tau_{\rm th}=\Rth\Cth$, swept via $\Cth$ alone at
fixed $\Rth$: both are insensitive to $\Cth$ while $\tau_{\rm th}$ stays far
below the forcing period ($66.7$~s, dotted line).}
\label{fig:Ea-sweep}
\end{figure}

This propagates into the switching dynamics. \Cref{fig:Ea-sweep}(a--c) scans
$\Ea\in[0,0.82]$~eV continuously at the baseline operating point: the
excursion grows from $0.712$ to $0.950$, the peak temperature from $302.8$ to
$307.0$~K, and the loop area by a factor $647$, all smoothly and monotonically
with no discontinuity anywhere in the range. That smooth, threshold-free
dependence is structural: ATDM's response to $\Ea$ is the continuous
exponential amplification of \cref{eq:arrhenius} acting on the drift rate,
with no combinatorial gate or switching threshold anywhere in the model.

Panels (d) and (e) sweep the thermal capacitance over thirteen orders of
magnitude above its material-derived value at fixed $\Rth$. \new{Both $T_{\max}$ and $\Delta x$ remain approximately
constant over the quasi-static portion of the sweep.
This behavior is consistent with the reduced thermal
relation $T\simeq\Tamb+\Rth P_{\rm Joule}$, which contains
no $\Cth$: independence from thermal capacitance holds
in the quasi-static limit. As the thermal relaxation time
becomes comparable to the timescale of the dissipated
power, this approximation breaks down; over the extended
sweep, $T_{\max}$ decreases from $306.2$ to $293.6$~K.} Reaching
that regime requires inflating $\Cth$ by ten to thirteen decades, so this
sweep is a stress test of the quasi-static assumption, not an exploration of
plausible values---and, as \cref{sec:collapse} notes, the weak influence of $\Cth$ \new{is
consistent with} its absence from $\Theta$.

\subsection{Regression against the classical HP model}
\label{sec:validation-ATDM}

\begin{figure}[ht!]
\centering
\includegraphics[width=8cm,height=7cm, trim = 0 0 0 0, clip]{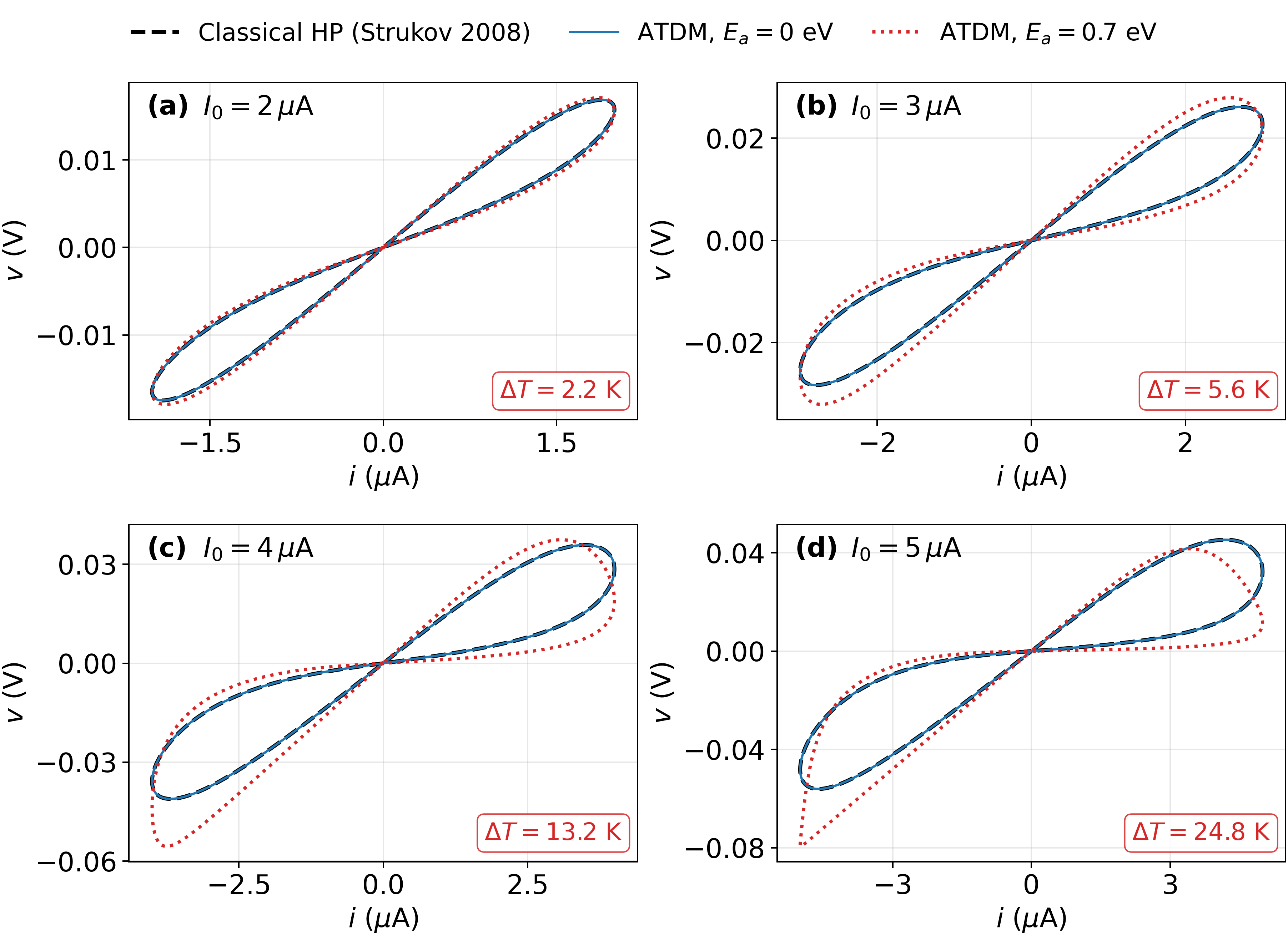}
\caption{Recovery of the isothermal limit across amplitude, at fixed
$f=0.015$~Hz: classical HP model (Strukov \textit{et al.}
\cite{Strukov2008}, dashed black), ATDM at $\Ea=0$~eV (blue), and ATDM at
$\Ea=0.7$~eV (dotted red), for (a) $I_0=2$, (b) $3$, (c) $4$, and
(d) $5\,\mu$A. The first two coincide to within solver tolerance in every
panel. The third departs by an amount that grows with $I_0$; the boxed value
reports the corresponding peak temperature rise above ambient.}
\label{fig:validationHP}
\end{figure}

\begin{figure}[ht!]
\centering
\includegraphics[width=8cm,height=7cm, trim = 0 0 0 0, clip]{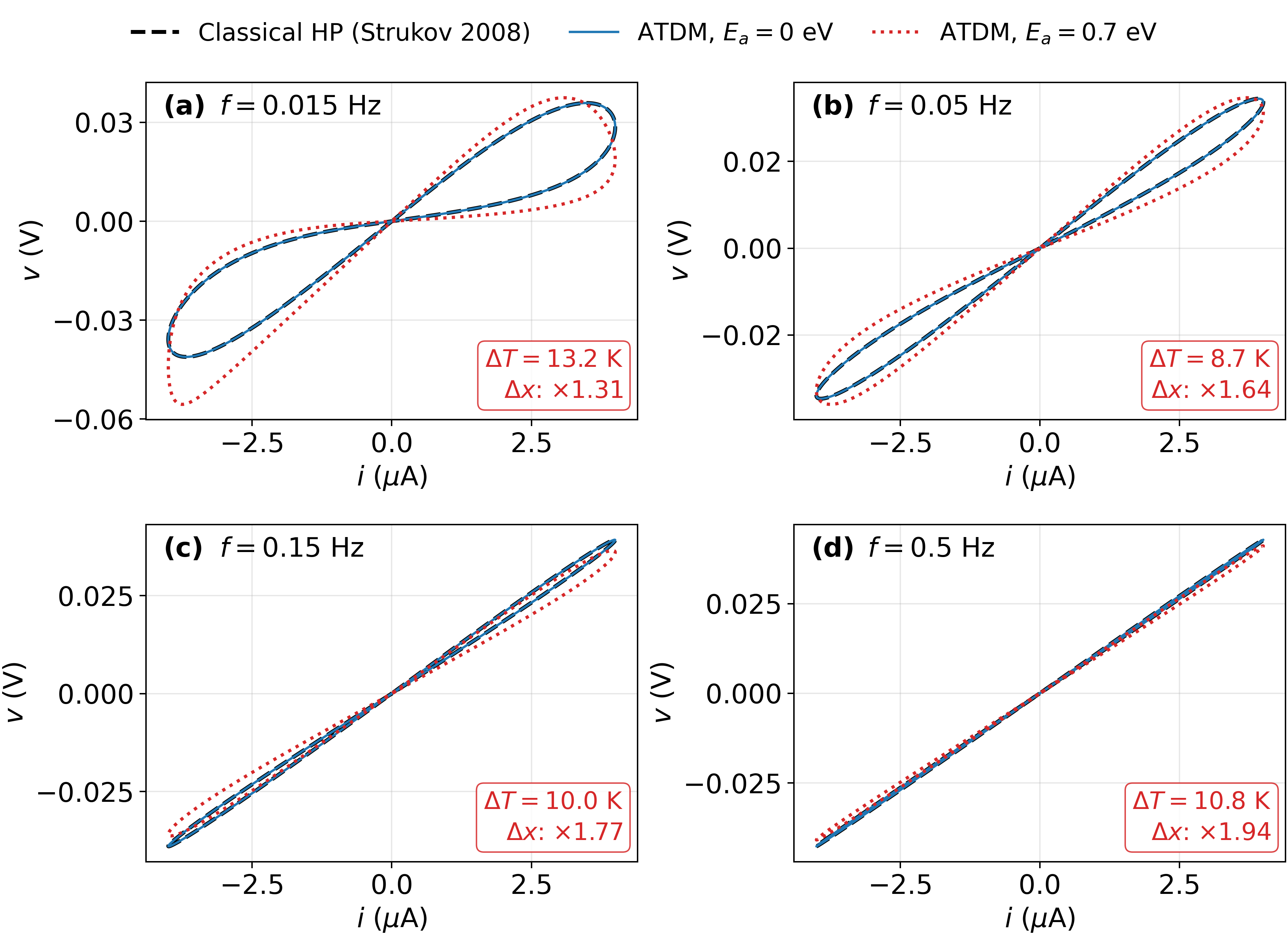}
\caption{The same test across frequency, at fixed $I_0=4\,\mu$A, for
(a) $f=0.015$~Hz, (b) $0.05$~Hz, (c) $0.15$~Hz and (d) $0.5$~Hz. The
coincidence of the first two curves holds at every frequency, to at most
$10^{-7}$~V. The boxed values give the peak temperature rise and the ratio of
state-variable excursions with and without thermal feedback: the relative
effect of the feedback grows from $1.31$ to $1.94$ as the loop itself
degenerates toward a straight line.}
\label{fig:validationHP-freq}
\end{figure}

ATDM is built to recover its reference model in the appropriate limit: as
$\Ea\to0$, $\muv(T)=\mu_0$ and \cref{eq:system3a} returns, term for term, to
the isothermal drift equation of Strukov \textit{et al.} \cite{Strukov2008}.
The test is run numerically rather than analytically, and along both axes of
the operating plane: four amplitudes at fixed frequency
(\cref{fig:validationHP}) and four frequencies at fixed amplitude
(\cref{fig:validationHP-freq}). Each panel compares (i) the classical model
implemented independently---constant $\mu_0$, no thermal equation at all, so
that the comparison is a genuine cross-check and not a tautology, (ii) ATDM at
$\Ea=0$, and (iii) ATDM at $\Ea=0.7$~eV.

\paragraph{Amplitude.} Curves (i) and (ii) coincide to
$10^{-8}$--$10^{-7}$~V in every panel of \cref{fig:validationHP}, about
$10^{-4}\,\%$ of the peak voltage---solver tolerance rather than a modeling
discrepancy, and the isothermal limit of \cref{sec:tr3a} verified rather than
asserted. Curve (iii) departs visibly, and the boxed value quantifies the
self-heating responsible: $\Delta T$ grows monotonically from $2.2$~K at
$I_0=2\,\mu$A to $5.6$, $13.2$, and $24.8$~K at $3$, $4$, and $5\,\mu$A. The
frequency $f=0.015$~Hz was selected empirically---the enclosed area is
maximal and stable near this value, the state variable traversing most of
$[0,1]$ within a half-cycle without saturating---and sits many orders of
magnitude below the $\varepsilon=1$ boundary of
\cref{sec:qs-frequency-range}.

\paragraph{Frequency.} \Cref{fig:validationHP-freq} repeats the test at fixed
$I_0=4\,\mu$A across the range used throughout this work. The coincidence
holds at every frequency, to at most $10^{-7}$~V
($3\times10^{-4}\,\%$ of the peak voltage): the isothermal limit is a property
of the equations and not a coincidence at one operating point. The thermal
feedback, by contrast, becomes relatively \emph{more} important as the
frequency rises. The ratio $\Delta x(0.7\,\mathrm{eV})/\Delta x(0)$ grows
monotonically from $1.31$ at $0.015$~Hz to $1.64$ at $0.05$~Hz, $1.77$ at
$0.15$~Hz and $1.94$ at $0.5$~Hz, while the absolute excursion collapses from
$0.94$ to $0.05$ and the loop degenerates toward the ohmic line of panel (d).
The reason is the timescale separation of \cref{sec:tau}: raising the
frequency leaves less time for ionic drift but none for the thermal response
to lag, so the peak rise stays within $8.7$--$13.2$~K across the whole sweep
while the excursion falls twentyfold, and a larger share of what excursion
remains is owed to the mobility enhancement. The two figures are therefore
complementary rather than redundant---the departure from the isothermal limit
grows with amplitude in absolute terms and with frequency in relative terms.

\begin{figure}[ht!]
\centering
\includegraphics[width=8.4cm,height=8cm, trim = 0 0 0 0, clip]{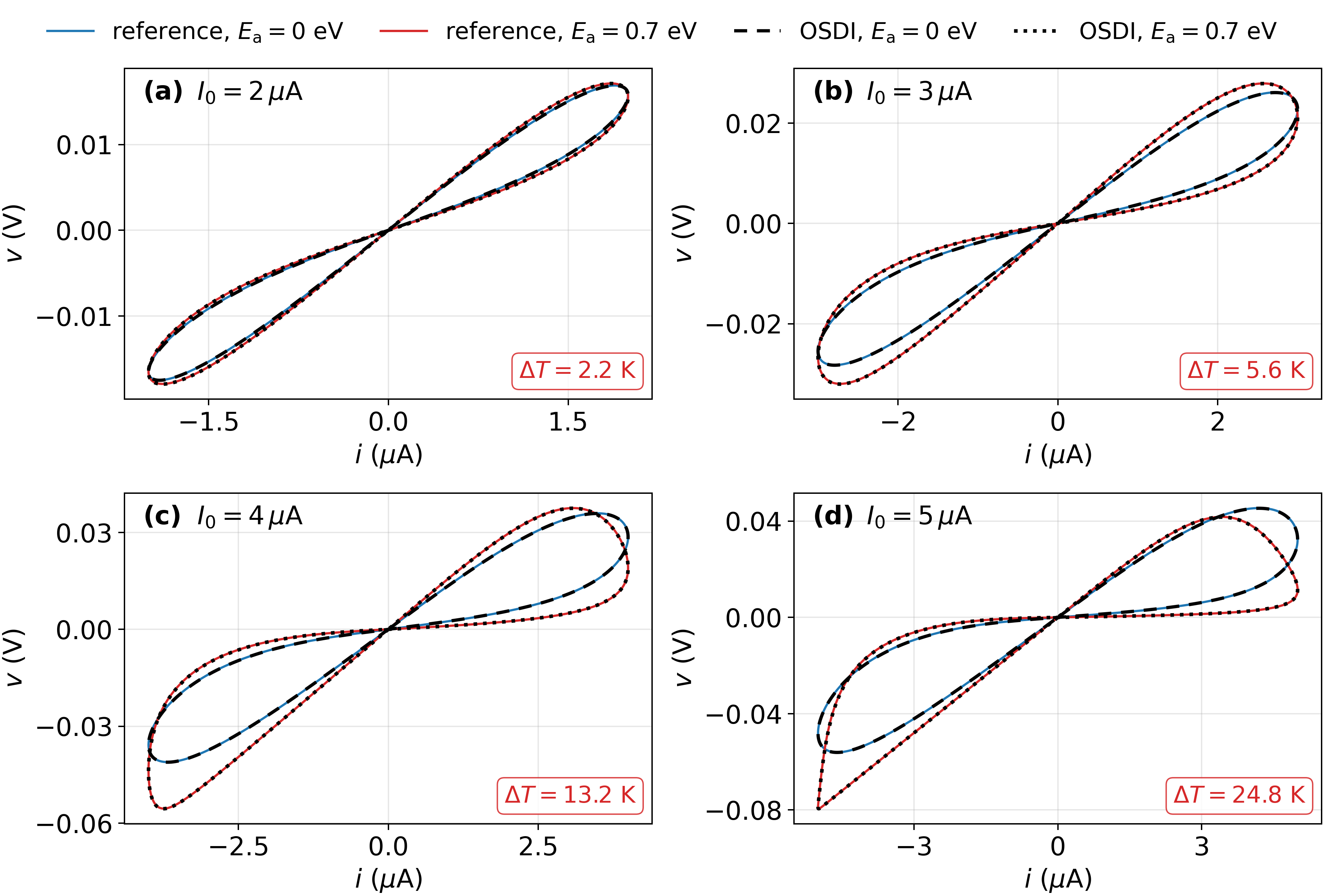}
\caption{\new{The same regression test as \cref{fig:validationHP}, rerun
through the \textsc{verilog-a} compact model of \cref{app:veriloga}, compiled
with OpenVAF and executed in \texttt{ngspice}~47: compiled model in black,
dashed at $\Ea=0$~eV and dotted at $\Ea=0.7$~eV, against the reference
transcription of the same equations in blue and red, at the same four
amplitudes. The two are indistinguishable at plotting resolution. The boxed
peak temperature rises reproduce those obtained with Radau, and at $\Ea=0$ the
compiled model returns the classical isothermal excursion.}}
\label{fig:veriloga}
\end{figure}

\paragraph{Independence of the integration engine.} Both tests above use the
same adaptive solver, so alone they cannot exclude a shared solver artifact.
\new{\Cref{fig:veriloga} repeats the amplitude test through the
\textsc{verilog-a} compact model of \cref{app:veriloga}, compiled to a shared
object with OpenVAF and run as a device inside \texttt{ngspice}~47. Nothing is
shared with the reference beyond the equations themselves: a different code
base, the two states carried as circuit nodes rather than as an ODE state
vector, and the simulator's own adaptive trapezoidal--Gear integration in
place of Radau. Over the settled period the compiled model reproduces the
reference to within $0.009\,\%$ of the peak voltage in the worst of the eight
cases and $0.002\,\%$ in seven of them, and returns the same peak rises \new{to within $0.02$~K},
$2.17$, $5.64$, $13.22$ and $24.79$~K. At $\Ea=0$ the Arrhenius factor is
identically unity, so the drift equation is the isothermal one term for term,
and the simulator then returns $x\in[0.1441,0.8561]$ over the settled period
at $I_0=4\,\mu$A, against $[0.144,0.856]$ for the classical model.}
Conformity with Strukov \textit{et al.} in the isothermal limit is a property
of the equations, reproduced \new{by an independent implementation running in
an independent simulator}.

\begin{figure*}[!tb]
\centering
\includegraphics[width=17cm,height=3.8cm, trim = 0 0 0 0, clip]{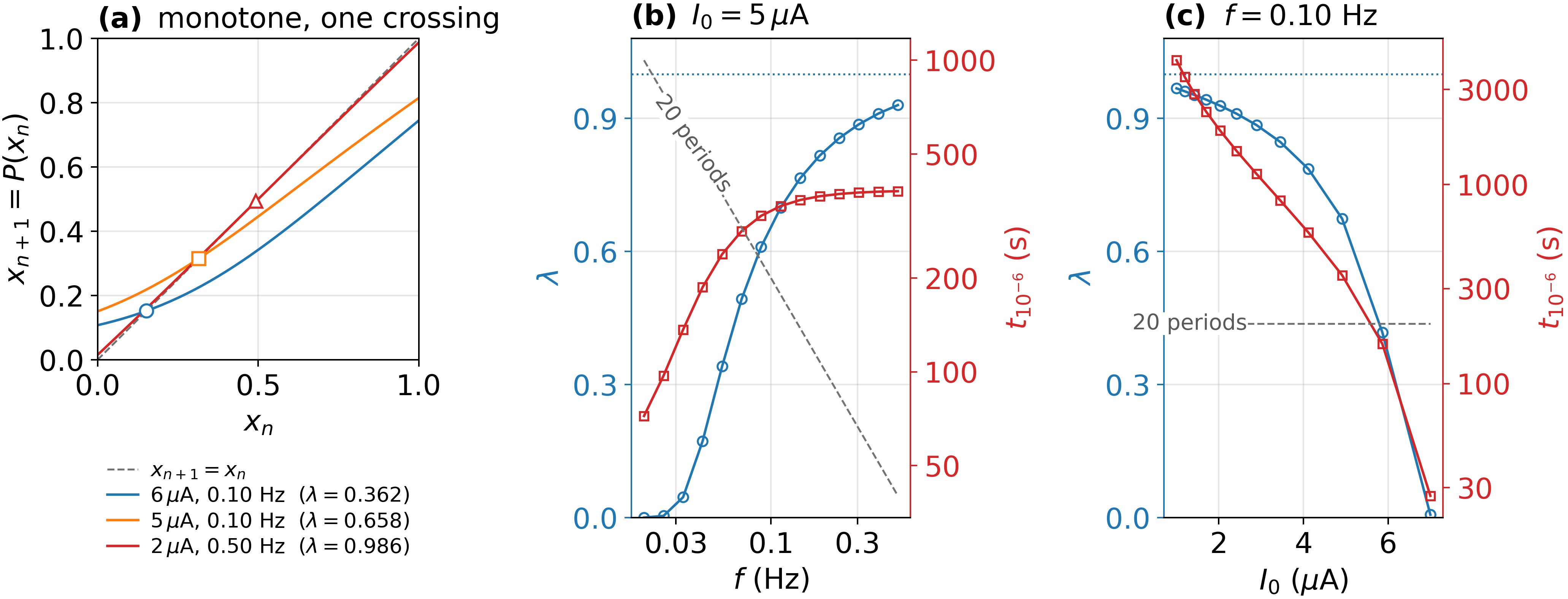}
\caption{Poincar\'e map of ATDM and its Floquet multiplier ($\Ea=0.7$~eV).
(a) One-period stroboscopic map $P:x_n\mapsto x_{n+1}$ at three operating
points, with the diagonal (dashed) and the fixed point of each (open
markers). $P$ is strictly increasing and crosses the diagonal exactly once, so
\new{it has no periodic orbits of period greater than one}. (b) Floquet
multiplier $\lambda=P'(x^*)$ (blue, left axis) and the settling time
$t_{10^{-6}}$ of \cref{eq:settling} (red, right axis) against frequency at
$I_0=5\,\mu$A; (c) the same against amplitude at $f=0.10$~Hz. The multiplier
stays strictly inside the unit circle over both sweeps and approaches unity as
the frequency rises or the current falls, which is what makes the relaxation
slow there. The dashed line is a $20$-period transient, $20/f$ seconds.}
\label{fig:poincare}
\end{figure*}

\subsection{\new{Monotonicity and period-one behavior in the quasi-static reduction}}
\label{sec:poincare}

To confirm that these pinched loops are genuine period-1 limit cycles rather
than longer-period or quasi-periodic responses, we work with the stroboscopic
map of the model: the state one forcing period later as a function of the
state now, $P:x_n\mapsto x_{n+1}=x(t_n+1/f)$. The argument below concerns the
\emph{reduced} system under current drive, in which the temperature is slaved
to the state; the step from the full two-state model to that reduction is
quantified after the statement. \new{The proposition excludes periodic orbits
of period greater than one for this reduction; it establishes neither the
uniqueness of the fixed point nor any property of the full two-state map, both
of which are addressed numerically below.}

\begin{proposition}[Reduced stroboscopic map]
\label{prop:monotone}
Fix $(I_0,f,\Ea,\Tamb)$ and let the temperature be slaved,
$T_{\rm QS}(t,x)=\Tamb+\Rth\,i(t)^2R(x)$, so that the dynamics reduce to the
scalar nonautonomous equation $\dot x=g(t,x)$ with
$g(t,x)=\muv(T_{\rm QS}(t,x))\Ron\,i(t)F(x,i)/D^{2}$, $g$ being $1/f$-periodic
in $t$. Assume \emph{(i)} $g$ is continuous in $t$ and locally Lipschitz in
$x$ on $[0,1]$, and \emph{(ii)} $g(t,0)\ge0\ge g(t,1)$ for all $t$. Then the
period map $P(x_0)=x(t_0+1/f;x_0)$ is a well-defined, continuous, and strictly
increasing map of $[0,1]$ into itself---a homeomorphism onto its image, not
necessarily onto $[0,1]$---it has at least one fixed point, and it has no
periodic orbit of period greater than one.
\end{proposition}

Assumption (i) holds because $R$, $\muv$, and $F$ are smooth in $x$ and the
drift term $iF(x,i)$ has the same zero limit on both sides of a current zero
crossing, so the piecewise definition of \cref{eq:biolek} extends continuously;
(ii) is what the window is for. The conclusions then follow from standard
facts: uniqueness of solutions forbids two trajectories from crossing, so
$P$ preserves order and is injective, and continuous dependence on initial
data makes it continuous, hence a homeomorphism onto its image $P([0,1])$;
(ii) makes $[0,1]$ forward invariant, and a continuous map of an interval into
itself has a fixed point; and if
$P$ is increasing then $P^n(x)>x$ for all $n$ whenever $P(x)>x$, so a periodic
point of period $n>1$ cannot exist \cite{Strogatz2024}. Period-doubling,
quasi-periodicity, cycles of period $\ge2$, and chaos are therefore excluded
for the reduced system---structurally, not as the outcome of a sweep. What
monotonicity does \emph{not} exclude is a plurality of coexisting attracting
fixed points, which an increasing map may perfectly well possess.

Two things the proposition does \emph{not} give should be stated plainly. It
does not give \emph{uniqueness} of the fixed point, only that every periodic
point is one; uniqueness is a numerical finding here, the map crossing the
diagonal exactly once at every operating point tested. And it does not by
itself cover the full two-state model, whose map acts on $(x,T)$. \new{Scale separation supports this reduction.
For a prescribed electrical trajectory, thermal
perturbations decay at the rate
$1/\tau_{\rm th}\approx5.5\times10^{9}~\mathrm{s}^{-1}$.
In the coupled system, a temperature perturbation can
also modify the electrical trajectory, so this decay
rate alone does not establish contraction of the full
two-state map.} Numerically, two
initial temperatures $80$~K apart return the same $x$ after one period to
within $4\times10^{-11}$. \new{This provides numerical support for applying
the reduced description to the full model under the tested quasi-static conditions; it is
not a proof that the proposition transfers to the two-state map.}

\Cref{fig:poincare}(a) shows $P$ at three operating points together with the
diagonal: monotone increasing, crossing it exactly once. \new{At these points
the fixed point $x^*$ is therefore numerically unique}, and its Floquet multiplier $\lambda=P'(x^*)$---real and positive,
since $P$ increases---measures how fast the orbit is approached:
$\lambda=0.362$, $0.658$ and $0.986$ at the three points. Across the two sweeps of
\cref{fig:poincare}(b,c)---frequency at $I_0=5\,\mu$A, amplitude at
$f=0.10$~Hz---$\lambda$ stays within $[1\times10^{-4},0.968]$. \new{Since
$0<\lambda<1$, the period-1 orbit is locally attracting at every point of
these two sweeps, namely $f\in[0.02,0.5]$~Hz at $I_0=5\,\mu$A and
$I_0\in[1,7]\,\mu$A at $f=0.10$~Hz; the multiplier does not establish
uniqueness, which rests separately on the single numerical intersection of $P$
with the diagonal at the points tested. The full $(I_0,f)$ rectangle has not
been swept.} The third operating point of panel (a),
$I_0=2\,\mu$A at $f=0.50$~Hz, lies on neither sweep and is the most weakly
contracting of the three, $\lambda=0.986$: it is shown precisely because it
approaches the marginal case without reaching it. A direct stroboscopic
sweep at $150$ points per axis agrees, as it must: once the transient has
decayed, $x$ collapses to a single value at every point of the grid, with a
residual spread $\lesssim2\times10^{-7}$.

That proviso deserves to be made quantitative, because a transient that has
not decayed is easily misread as a bifurcation---and the multiplier is
exactly what quantifies it. \new{Close to the periodic orbit, where the
linearization holds, a deviation is multiplied by $\lambda$ at each period, so
an asymptotic estimate of the time needed to reduce it by a factor $\delta$ is}
\begin{equation}
t_\delta=\frac{\ln(1/\delta)}{f\,\lvert\ln\lambda\rvert}\quad\text{seconds},
\label{eq:settling}
\end{equation}
an \emph{elapsed} time, not a period count. Since $\lambda\to1$ as the
frequency rises or the current falls, $t_\delta$ grows at both ends of the
campaign: at $\delta=10^{-6}$ it reaches $380$~s at $f=0.5$~Hz
($I_0=5\,\mu$A) and $4.2\times10^{3}$~s at $I_0=1\,\mu$A ($f=0.10$~Hz). The
dashed lines in \cref{fig:poincare}(b,c) mark a fixed $20$-period transient,
that is $20/f$ seconds: it falls short of $t_{10^{-6}}$ above
$f\approx0.09$~Hz and below $I_0\approx4.9\,\mu$A---exactly the two regions
where an earlier exploratory pass showed residual scatter, a spread of
$4.3\times10^{-2}$ at $I_0=1.2\,\mu$A along the current sweep and
$2.4\times10^{-2}$ at $0.50$~Hz along the frequency sweep, against
$7.0\times10^{-8}$ at $0.05$~Hz where $20$ periods already amount to $400$~s.
Recomputing with the transient set by elapsed time confirms the diagnosis:
discarding $500$ periods brings the spread below $2\times10^{-7}$ at every
point tested. The trap is generic to stroboscopic studies of drift-based
memristor models, since the approach to the periodic orbit is governed by the
drift rate $\muv\Ron I_0/D^2$ and hence by a physical time---of order
$300$~s at $I_0=5\,\mu$A, scaling as $1/I_0$---whereas a transient counted in
periods lasts $n/f$ seconds and shortens as the frequency rises.

How much of this depends on the window is worth separating.
\Cref{app:robustness} repeats the construction with the Joglekar window
$F=1-(2x-1)^{2p}$, which does not depend on the sign of the current.
Monotonicity survives, as \cref{prop:monotone} requires---its hypotheses hold
for either window---and with it the exclusion of period-doubling and chaos.
The \emph{contraction} does not: the Joglekar map is the identity to within
$10^{-6}$, every state is a neutral fixed point, and the settled orbit keeps
the memory of its initial condition instead of selecting one. Attractor
selection in ATDM is therefore a property of the sign-switching window, while
the exclusion of complex dynamics is a property of the scalar reduction. The
two claims have different scopes and are stated separately here for that
reason.

\subsection{\new{Absence of a ratchet regime in the quasi-static reduction}}
\label{sec:ratchet}

The same settled-orbit analysis answers a question raised by filamentary
electrothermal models. In Pickett-type formulations
\cite{Pickett2009,PickettThesis} the switching rate is exponential in the
applied voltage, so below a drive threshold the tunneling gap can advance in
one direction without returning within a cycle---a ratchet regime, separated
from bounded oscillation by a sharp amplitude threshold.

\begin{figure*}[ht!]
\centering
\includegraphics[width=16cm,height=4cm, trim = 0 0 0 0, clip]{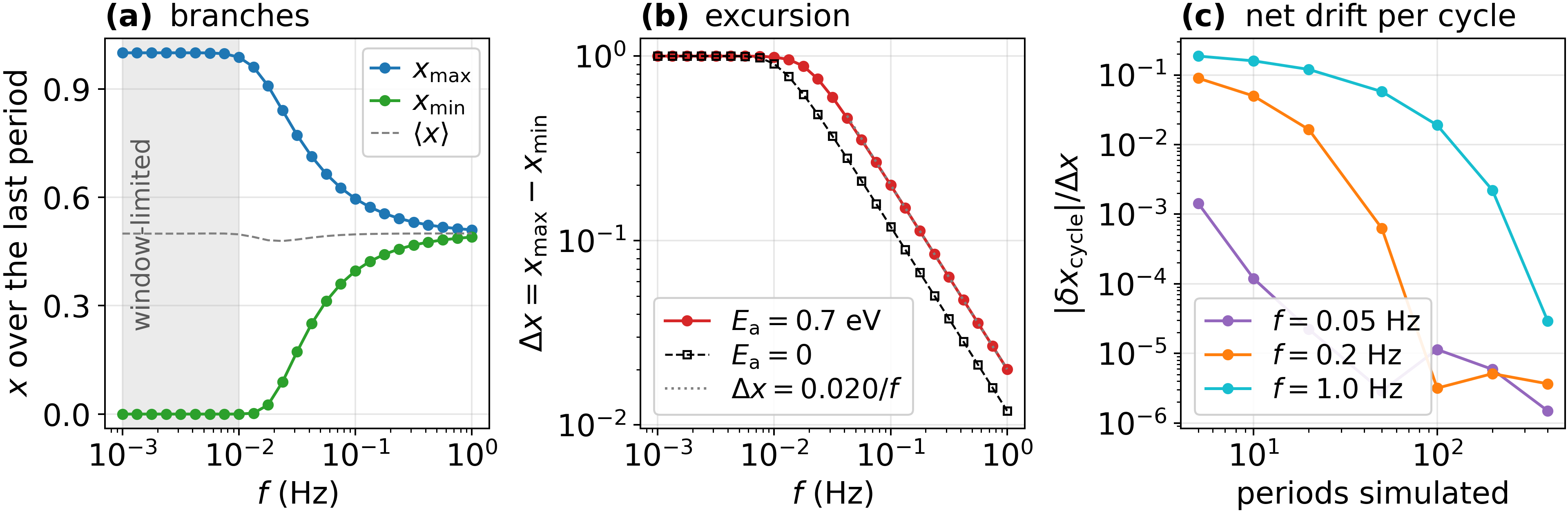}
\caption{ATDM exhibits no ratchet regime. (a) Extrema $x_{\min}$ and
$x_{\max}$ of the state variable over the last forcing period, and their mean,
against frequency at $I_0=4\,\mu$A, $\Ea=0.7$~eV; the branches close
continuously and the shaded band marks the low-frequency region where the
window pins the state at its bounds. (b) The excursion follows
$\Delta x=0.020/f$ to within $0.8\,\%$ over the twelve points of the linear
regime, at $\Ea=0.7$~eV and in the isothermal limit alike, the activation
energy contributing only the constant factor $1.68$. (c) Net displacement over
one cycle, normalized by the excursion, against the number of periods
simulated: it decays toward zero at every frequency, so the apparent drift
seen over a few periods is a relaxation transient and not a sustained
ratchet.}
\label{fig:ratchet}
\end{figure*}

ATDM admits no such regime. \Cref{fig:ratchet} sweeps the frequency over three
decades at $I_0=4\,\mu$A: the branches $x_{\min}$ and $x_{\max}$ close
continuously with no threshold, the settled orbit is symmetric
($\langle x\rangle=0.500$ wherever the window does not pin the state at its
bounds), and the excursion follows $\Delta x=0.020/f$ to within $0.8\,\%$
across the linear regime---the charge injected per half-cycle, referred to
the film thickness. Within that linear regime the activation energy enters the
amplitude only as a constant factor $1.68$, not as a threshold; outside it,
where the window pins the state or the self-heating couples strongly to $x$,
no such constant factor is claimed. The symmetry $\langle x\rangle=1/2$ is
likewise a property of the sign-switching Biolek window under a symmetric
drive rather than of the drift structure alone: \cref{app:robustness} shows
that it does not survive a window independent of the current sign. \new{For the quasi-statically reduced model, the absence of
sustained net drift follows from the stroboscopic-map structure
established in \cref{sec:poincare}. The period map is continuous
and increasing on $[0,1]$, so its iterates are monotone and
bounded and converge to a fixed point. Consequently, the net
state displacement per forcing period tends to zero, and the
signed displacements over the two half-cycles cancel on the
limiting period-1 orbit. This does not exclude transient drift
during relaxation.} A rate
exponential in the drive, by contrast, can be negligible on one half-cycle and
finite on the other, which is what locks a filamentary model on one side.
\Cref{fig:ratchet}(c) closes the argument in a different variable: the net
displacement per cycle decays toward zero as the simulation lengthens, at
every frequency tested.

\subsection{Temperature-dependent hysteresis}
\label{sec:hysteresis-vs-Tamb-3a}

\begin{figure}[ht!]
\centering
\includegraphics[width=8cm,height=3.5cm, trim = 0 0 0 0, clip]{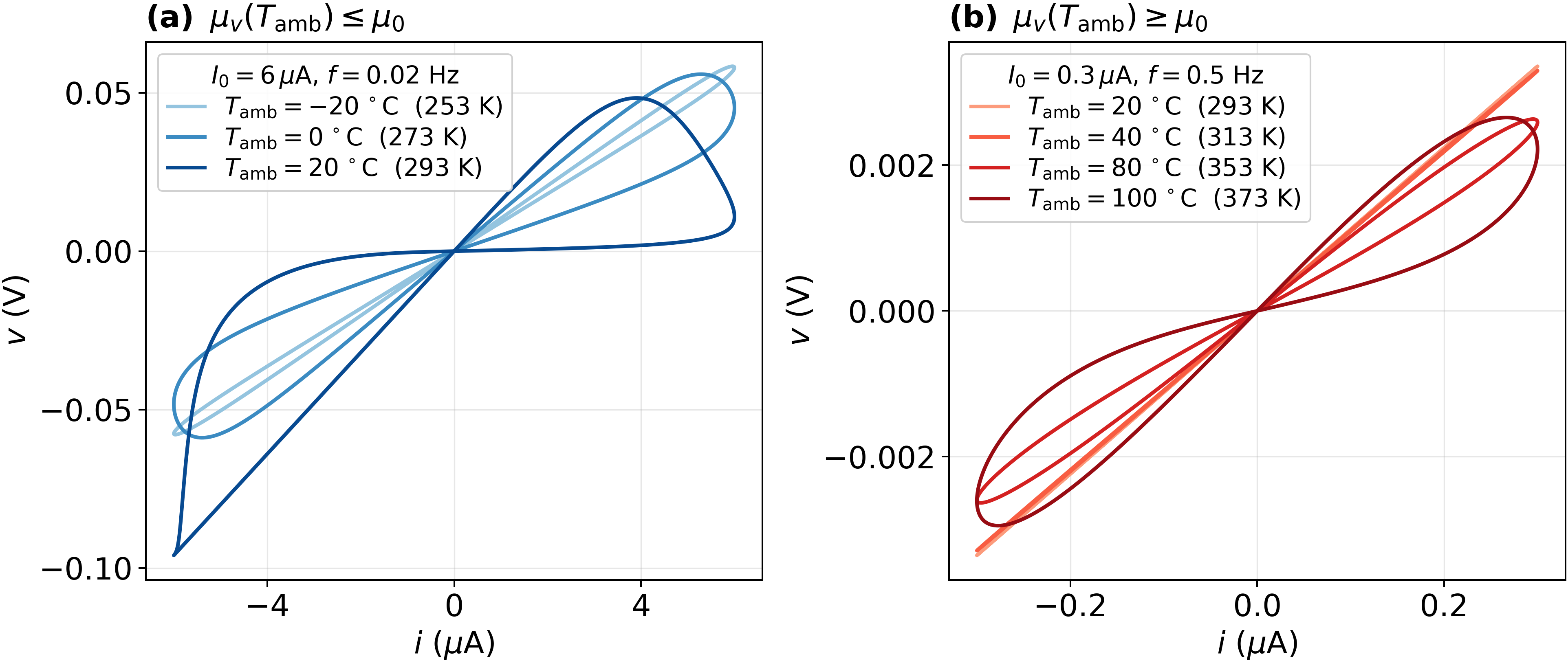}
\caption{Pinched current--voltage loops against ambient temperature,
$\Ea=0.7$~eV, $T_0=293$~K in both panels. (a) $\muv(\Tamb)\le\mu_0$
($\Tamb=-20,0,20\,^\circ$C, $I_0=6\,\mu$A, $f=0.02$~Hz). (b)
$\muv(\Tamb)\ge\mu_0$ ($\Tamb=20,40,80,100\,^\circ$C, $I_0=0.3\,\mu$A,
$f=0.5$~Hz). Each panel's operating point is given as the title of its legend.
The two differ because $\muv(\Tamb)$ varies by some four orders of magnitude
over this range; no single $(I_0,f)$ avoids saturation throughout.}
\label{fig:hysteresis-tamb}
\end{figure}

Since $\muv(\Tamb)$ varies by $\sim3\times10^4$ between $-20$ and
$100\,^\circ$C at $\Ea=0.7$~eV, a single operating point tuned for one end of
that range collapses to a near-flat loop at the other, or saturates into a
degenerate line. \Cref{fig:hysteresis-tamb} therefore uses two: one for
$\muv(\Tamb)\le\mu_0$ (panel a, $I_0=6\,\mu$A, $f=0.02$~Hz) and one for
$\muv(\Tamb)\ge\mu_0$ (panel b, $I_0=0.3\,\mu$A, $f=0.5$~Hz). Panel (a)
widens monotonically from $-20$ to $20\,^\circ$C; panel (b) shares
$20\,^\circ$C as reference---nearly flat there, since it is tuned for the far
larger mobilities at $80$--$100\,^\circ$C, which illustrates concretely why no
single operating point serves both regimes---then widens again from $40$ to
$100\,^\circ$C without saturating.

\section{Quasi-static model order reduction}
\label{sec:qs-reduction}

Replacing the thermal equation by \cref{eq:quasistatic} whenever
$\varepsilon\ll1$ is a model order reduction: the state shrinks from $[x,T]$
to $[x]$, obtained not by a generic procedure---balanced truncation, proper
orthogonal decomposition---but by exploiting ATDM's intrinsic timescale
separation. Retaining the full two-state system makes that limit a derivable
consequence rather than an assumption. Since the drift dynamics contain no
combinatorial gate comparable to the exponential lock of filament-based
formulations, the reduced system is genuinely nonstiff and admits an explicit
solver (RK45) where the full system requires an implicit one.

\begin{figure}[ht!]
\centering
\includegraphics[width=8.7cm,height=4cm, trim = 0 0 0 1.7cm, clip]{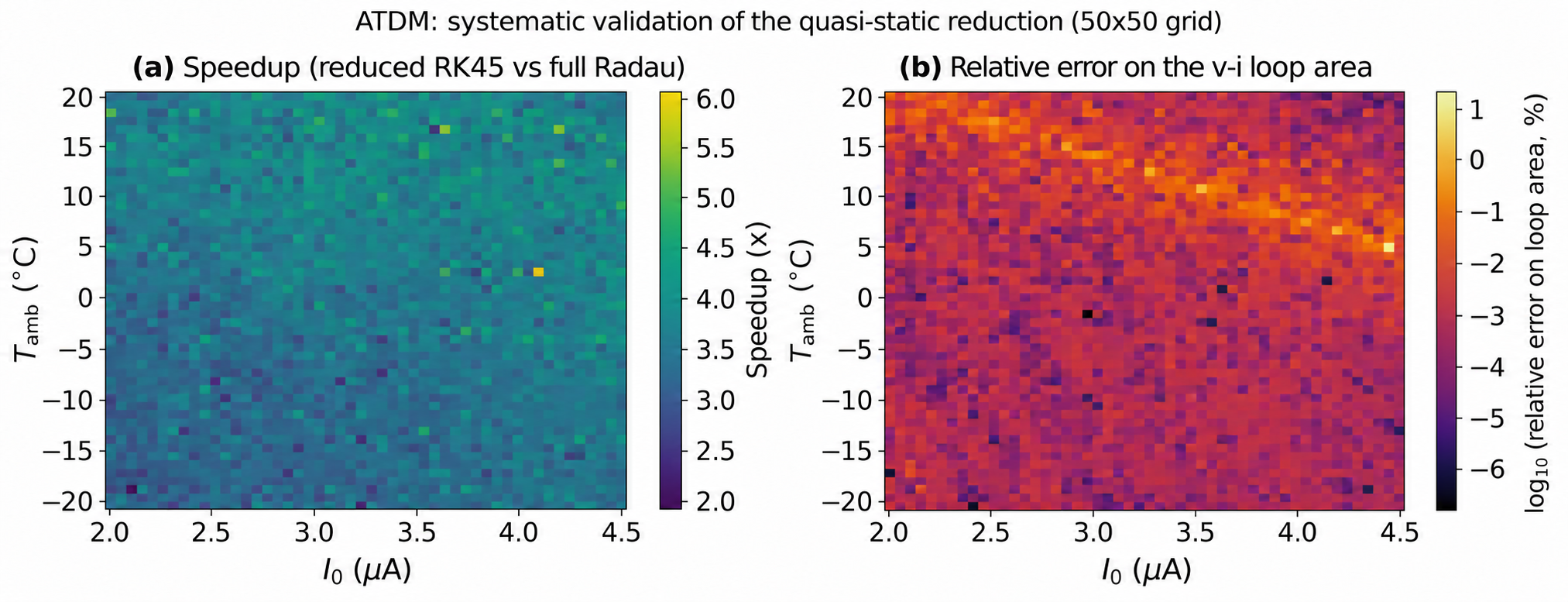}
\caption{Numerical verification of the quasi-static reduction over a
$50\times50$ grid in $(I_0,\Tamb)$. (a) Speedup of the reduced model (RK45,
$\mathrm{rtol}=10^{-9}$, $\mathrm{atol}=10^{-12}$) relative to the full model
(Radau, $\mathrm{rtol}=10^{-6}$, $\mathrm{atol}=10^{-9}$). (b) Relative error
on the pinched $v$--$i$ loop area, logarithmic color scale. The speedup stays
between $1.9\times$ and $6.1\times$; the error is at or below $0.1\,\%$ for
$99\,\%$ of the grid, with a single near-degenerate-loop outlier discussed in
the text.}
\label{fig:qs-speedup}
\end{figure}

\Cref{fig:qs-speedup} reports the speedup and the relative error on the
pinched-loop area over a $50\times50$ grid in $(I_0,\Tamb)$, both solvers at
matched, tightened tolerance. The speedup is $3.5\times$ on average
($1.9$--$6.1\times$), with the error at or below $0.1\,\%$ for $99\,\%$ of the
$2500$ points and a median of $0.001\,\%$. An exploratory pass at the reduced
model's own default tolerance had suggested a considerably larger speedup
($\sim13\times$ on a coarser grid), but that proved a tolerance artifact:
isolated points showed errors up to $5.5\,\%$ which vanished once the
tolerance was tightened. The $3.5\times$ figure is the defensible one.

A single point of the $2500$ shows a markedly larger error ($16.2\,\%$ at
$I_0=4.45\,\mu$A, $\Tamb\approx5.2\,^\circ$C), within a faint diagonal band of
mildly elevated error. This is a metric artifact: the loop there is nearly
degenerate---area $\sim2\times10^{-13}$, five orders of magnitude below
typical---so the relative-area metric is ill-conditioned, while the
trajectories still agree to $3\times10^{-7}$ on $x(t)$ and $0.0001\,\%$ on
$v(t)$. The same ill-conditioning reappears statistically in
\cref{sec:sobol}, and is resolved there by working on $\log_{10}$(area).

\new{A compiled \textsc{verilog-a} implementation of the full model, verified
against the reference inside a general-purpose simulator, is given in
\cref{app:veriloga}.} Taken together, the grid verification and \new{that
cross-check} support a bounded claim: within the baseline parameter ranges, the reduction reproduces the full
calculation with subpercent error, delivers a consistent speedup, and removes
the fast thermal stiffness. No equivalent claim is made for Pickett-type
formulations, whose exponentially gated kinetics impose their own
$T$-independent stiffness.

\section{Comparison with existing compact models}
\label{sec:comparison}

\Cref{tab:comparison} situates ATDM along five dimensions against the Strukov
drift model \cite{Strukov2008} and its window variant \cite{Biolek2009}, the
threshold-adaptive TEAM/VTEAM family \cite{Kvatinsky2013,Kvatinsky2015},
enhanced PSpice transcriptions of the \ce{TiO2} drift model
\cite{Kolka2015Enhanced}, and Pickett-type filament/tunneling-gap models
\cite{Pickett2009}, representative of the electrothermal treatments surveyed
in \cite{Kim2012,Mickel2014,Ielmini2011}.

``Minimal'' is meant in a checkable sense, and the criteria are these: \new{one
added thermal state} and no more; three added parameters ($\Rth$, $\Cth$, $\Ea$), of
which only two enter the quasi-static regime; the constitutive relation
$v=iR(x)$ unchanged; no switching threshold and no new gating function; \new{the
drift term bounded by the same window}; and the isothermal
model recovered exactly, term for term, as $\Ea\to0$. Each row of
\cref{tab:comparison} can be read against that list. Temperature-dependent
mobility appears in drift models before this work, lumped thermal networks
appear in filamentary and in RRAM compact models, and dimensionless switching
numbers appear in the scaling literature; \new{none of these ingredients is
claimed as new. The contribution claimed here is narrower and can be checked
item by item: (a)~the closed electrothermal loop is formulated within the
Strukov--Biolek drift structure under the six constraints above, so that the
isothermal model is recovered term for term; (b)~the response of that model is
expressed through the groups $\Pi$, $\Lambda$, $\langle\Gamma\rangle$, and
$\Theta$, with a closed-form approximation for $\langle\Gamma\rangle$, an
empirical collapse of the excursion, and a regime map in $(\varepsilon,\Theta)$
that quantifies the error of the quasi-static reduction; (c)~the stroboscopic
map of that reduction is proved to be monotone, which restricts its possible
periodic dynamics; and (d)~\new{the ATDM model has been implemented and verified as a \textsc{Verilog-A} compact device running in a general-purpose circuit simulator}. To our knowledge, no earlier drift-type compact model combines (a)--(d).}
The closest precedent within the drift family is the
temperature-dependent analytical treatment of Singh and Raj \cite{Singh2018},
which makes the model parameters functions of temperature but keeps the
temperature itself an externally imposed quantity; here it is a state, driven
by the dissipation the device produces, so the loop closes and the resulting
excursion becomes something that can be quantified and scaled.

A designer who needs only a fast threshold-switching macromodel, with no
interest in self-heating, is better served by TEAM/VTEAM or by generalized
behavioral formulations such as Yakopcic \textit{et al.}'s
\cite{Yakopcic2013}, both lighter than ATDM precisely because they carry no
thermal state. A researcher who needs the most detailed picture of filament
geometry and tunneling conduction is better served by a Pickett-type model.
ATDM targets a bounded middle ground: an assumed electrothermal feedback with
the computational simplicity of the drift family, without filament geometry or
an explicit threshold.

\begin{table*}[!tb]
\centering
\footnotesize
\caption{Comparison of ATDM with representative compact memristor models.}
\label{tab:comparison}

\begin{tabular}{@{}lcccc>{\raggedright\arraybackslash}p{0.36\textwidth}@{}}
\toprule
Model
& State
& Thermo--Dyn.
& $\mu_v(T)$
& Complexity
& What ATDM adds to it \\
\midrule

Strukov/HP \cite{Strukov2008}
& $x$
& No
& No
& Low
& Thermal state and $\muv(T)$, recovered exactly as $\Ea\to0$ \\

Biolek HP \cite{Biolek2009,Kolka2015Enhanced}
& $x$
& No
& No
& Low
& Same window, plus the feedback and its scaling groups \\

TEAM/VTEAM \cite{Kvatinsky2013,Kvatinsky2015}
& $x$
& No
& No
& Low--medium
& Self-heating without introducing a threshold \\

Pickett-type \cite{Pickett2009}
& Filament
& Often
& Yes
& High
& \new{Same feedback with one electrical and one thermal state; no filament geometry} \\

\textbf{ATDM (this work)}
& $x,T$
& \textbf{Yes}
& \textbf{Yes}
& Medium / low 
& \new{No switching threshold or filament physics} \\

\bottomrule
\end{tabular}
\end{table*}

The structural difference with Pickett-type models is not one of degree. These
conduct through a tunneling gap whose width is the state variable, so the
current--voltage relation is exponential in that width and extending them
means rederiving tunneling and gap-growth kinetics together. ATDM's loop is
added to the same minimal state and the same linear relation $v=iR(x)$:
temperature enters only through $\muv(T)$ multiplying the existing drift rate,
\new{with no new electrical state}, threshold, or output equation. That is why the extension is
minimal, and why the isothermal recovery of \cref{fig:validationHP} is exact.
\new{The same difference has a dynamical consequence: a rate exponential in the
drive can lock the state on one side, whereas for the quasi-statically reduced
ATDM the monotone stroboscopic map forces the net displacement per period to
vanish on the limiting orbit (\cref{sec:ratchet}). The absence of a ratchet
regime is therefore a property of the reduced drift structure studied here,
not a generic feature of electrothermal compact models.}

\section{Global sensitivity analysis}
\label{sec:sobol}

The sweeps of \cref{sec:numresults} vary one parameter at a time and cannot
apportion output variance between direct effects and interactions---the
coupled behavior an electrothermal model is built to capture. This section
closes that gap with a variance-based analysis (Sobol' \cite{Sobol2001} via
Saltelli sampling \cite{Saltelli2010}) over $(I_0,\Tamb,\Ea,\Rth)$, with
$\Rth\in[0.5,2]\times$ baseline---a plausible geometric and material
uncertainty, narrower than the deliberately extreme range of
\cref{sec:thermal-interpretation}. The design is $N=1024$ base points,
$N(D+2)=6144$ model evaluations, each a $30$-period simulation at $f=0.02$~Hz
from which three outputs are extracted: the excursion $\Delta x$, the peak
temperature $T_{\max}$, and the pinched-loop area, with
$\Delta T_{\max}=T_{\max}-\Tamb$ tracked separately to remove the additive
ambient term. \Cref{app:sobol} gives the estimators, the fixed-step
integration engine used to make a sample of this size tractable, and its
verification against the Radau reference.

\subsection{Rankings}
\label{sec:sobol-results}

\Cref{tab:sobol-4param} reports first-order, total-order, and interaction
indices. Three findings stand out.

First, $\Tamb$ dominates the excursion ($S_{T_i}=0.770$ against $0.242$ for
$I_0$) with a modest interaction term, and dominates the raw peak temperature
almost by construction: $T_{\max}\approx\Tamb+\Rth P_{\rm Joule}$, and $\Tamb$
is swept over $40$~K while the self-heating term spans only $2$--$16$~K.
Isolating $\Delta T_{\max}$ removes that artifact and reverses the ranking---
self-heating is governed by $I_0$ ($S_{T_i}=0.645$) and $\Rth$
($S_{T_i}=0.430$) jointly, exactly as $\Delta T_{\rm QS}=\Rth P_{\rm Joule}$
predicts.

Second, the loop area is governed overwhelmingly by \emph{interaction} rather
than by any parameter acting alone: $I_0$ and $\Tamb$ carry the largest total
effects ($0.767$ and $0.744$) of which $0.53$ and $0.61$ are interaction,
while $\Ea$ and $\Rth$ contribute almost exclusively through coupling
($S_i=0.003$ and $0.011$ against $S_{T_i}=0.226$ and $0.334$). Univariate
sweeps confirm this physically: the loop area grows with $I_0$ by a factor
$\sim550$ at $\Ea=0.19$~eV and $\sim1270$ at $0.82$~eV, so the activation
energy does not drive the area on its own so much as set how strongly the area
responds to the drive.

Third, and this revises a conclusion that the raw indices would support, the
loop-area indices are not resolved at this sample size. Over the ensemble the
raw area spans $7.5\times10^{-16}$ to $2.5\times10^{-8}$~V$\cdot$A---a ratio
of $3.4\times10^{7}$---with skewness $8.1$, and the largest $1\,\%$ of points
carry $36\,\%$ of the total. Sobol' indices are variance estimators, so every
index built on that sample is controlled by a handful of realizations: the
$95\,\%$ bootstrap intervals for $S_{T_i}(I_0)$ and $S_{T_i}(\Tamb)$ coincide
and their upper bounds exceed unity, which a total-order index cannot do. This
is the ill-conditioning already met on the near-degenerate loops of
\cref{sec:qs-reduction}, seen from the statistical side. Decomposing instead
the variance of $\log_{10}$(area)---the natural measure of dispersion for a
quantity spanning ten decades---resolves it at no computational cost, and
reverses one ranking: on the conditioned metric $\Ea=0.262\,[0.225,0.307]$
outranks $\Rth=0.145\,[0.127,0.167]$, whereas the raw metric had them the
other way round ($0.226$ against $0.334$).

The reversal holds at every $N$ from $64$ upward and in both the three- and four-parameter ensembles, so it is
a property of the metric rather than of the sample---and, by the same token,
a ranking \emph{conditional on the metric} rather than an absolute physical
one, the choice between the two being made on conditioning grounds.
Convergence supports the same reading: reestimating for
$N\in\{64,128,256,512,1024\}$ narrows the interval on $S_{T_i}(I_0)$ for the
loop area from $3.55$ to $0.70$ and leaves the ranking unchanged from $N=256$
onward, so ranking stability is established where index resolution is not.
\Cref{app:sobol} reports the intervals, that convergence and the two metrics
side by side; the conclusion on $\Delta T_{\max}$ is unaffected, $\Rth$ remaining a first-order contributor there, $S_{T_i}=0.430\,[0.377,0.489]$.

Even on the conditioned metric $I_0$ $[0.604,0.763]$ and $\Tamb$
$[0.557,0.696]$ still overlap. The defensible statement is that the two
jointly dominate the loop area without being separable at this sample size.

\begin{table}[ht!]
\centering
\footnotesize
\caption{Sobol' first-order ($S_i$), total-order ($S_{T_i}$), and interaction
indices over $(I_0,\Tamb,\Ea,\Rth)$; $N=1024$, $6144$ evaluations,
$n_{\rm periods}=30$, $f=0.02$~Hz. $\Delta T_{\max}=T_{\max}-\Tamb$ isolates
self-heating from the additive ambient term.}
\label{tab:sobol-4param}
\begin{tabular}{llccc}
\toprule
Output & Parameter & $S_i$ & $S_{T_i}$ & $S_{T_i}-S_i$\\
\midrule
\multirow{4}{*}{$\Delta x$}
 & $I_0$   & $0.162$ & $0.242$ & $0.080$ \\
 & $\Tamb$ & $0.667$ & $0.770$ & $0.103$ \\
 & $\Ea$   & $0.039$ & $0.074$ & $0.035$ \\
 & $\Rth$  & $0.018$ & $0.042$ & $0.023$ \\
\midrule
\multirow{4}{*}{$T_{\max}$}
 & $I_0$   & $0.098$ & $0.088$ & $-0.009$ \\
 & $\Tamb$ & $0.868$ & $0.870$ & $0.003$ \\
 & $\Ea$   & $0.004$ & $0.001$ & $-0.003$ \\
 & $\Rth$  & $0.037$ & $0.059$ & $0.022$ \\
\midrule
\multirow{4}{*}{$\Delta T_{\max}$}
 & $I_0$   & $0.549$ & $0.645$ & $0.096$ \\
 & $\Tamb$ & $0.015$ & $0.050$ & $0.036$ \\
 & $\Ea$   & $0.003$ & $0.009$ & $0.006$ \\
 & $\Rth$  & $0.328$ & $0.430$ & $0.102$ \\
\midrule
\multirow{4}{*}{Loop area}
 & $I_0$   & $0.236$ & $0.767$ & $0.531$ \\
 & $\Tamb$ & $0.139$ & $0.744$ & $0.606$ \\
 & $\Ea$   & $0.003$ & $0.226$ & $0.222$ \\
 & $\Rth$  & $0.011$ & $0.334$ & $0.323$ \\
\bottomrule
\end{tabular}
\end{table}

\subsection{Response maps and response surface}
\label{sec:sobol-maps}

\begin{figure*}[!bt]
\centering
\includegraphics[width=17.5cm,height=4cm, trim = 0 0 0 0, clip]{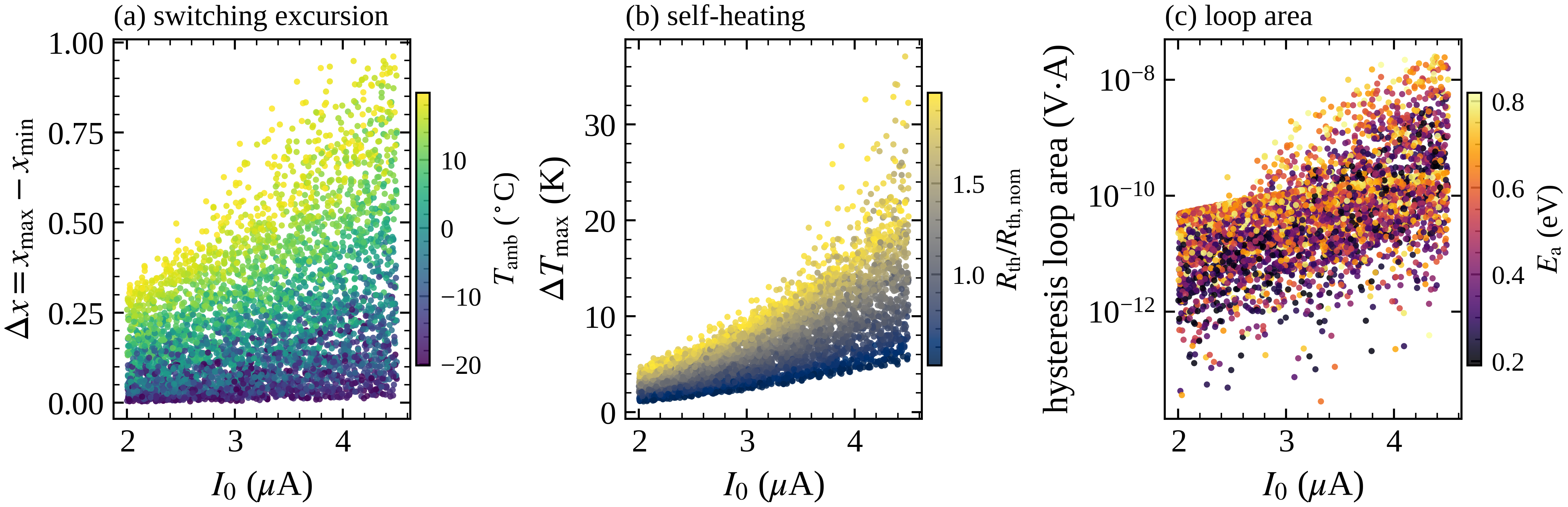}
\caption{Parameter-response maps from the $N=1024$, $D=4$ Saltelli campaign
($6144$ evaluations), each output projected onto the drive amplitude and
colored by a second parameter. (a) Excursion $\Delta x$, colored by $\Tamb$.
(b) Self-heating $\Delta T_{\max}$, colored by $\Rth/R_{\rm th,nom}$.
(c) Hysteresis loop area on a logarithmic ordinate, colored by $\Ea$. }
\label{fig:sobol-maps}
\end{figure*}

\Cref{fig:sobol-maps} gives the spatial counterpart of the variance ranking,
projecting the same ensemble onto the drive amplitude and coloring by a
second parameter. Panel (a) makes the dominance of $\Tamb$ over $\Delta x$
immediate: the sample separates into clean ambient bands, each rising gently
with $I_0$---the visual form of $S_{T_i}(\Tamb)=0.770$ against $0.242$.
Panel (b) shows self-heating controlled jointly by drive and thermal
resistance, the two entering as an almost separable product, consistent with
the small interaction terms for that output. Panel (c), on the logarithmic
ordinate the conditioning argument requires, shows the loop area stratified by
$\Ea$: at fixed $I_0$ the median area is $3.7$ times larger for $\Ea>0.7$~eV
than for $\Ea<0.3$~eV, with a spread exceeding two decades---the
interaction-dominated character of this output made visible. Four
thousand of the $6144$ points are drawn per panel.

\begin{figure*}[!bt]
\centering
\includegraphics[width=17.5cm,height=4cm, trim = 0 0 0 0, clip]{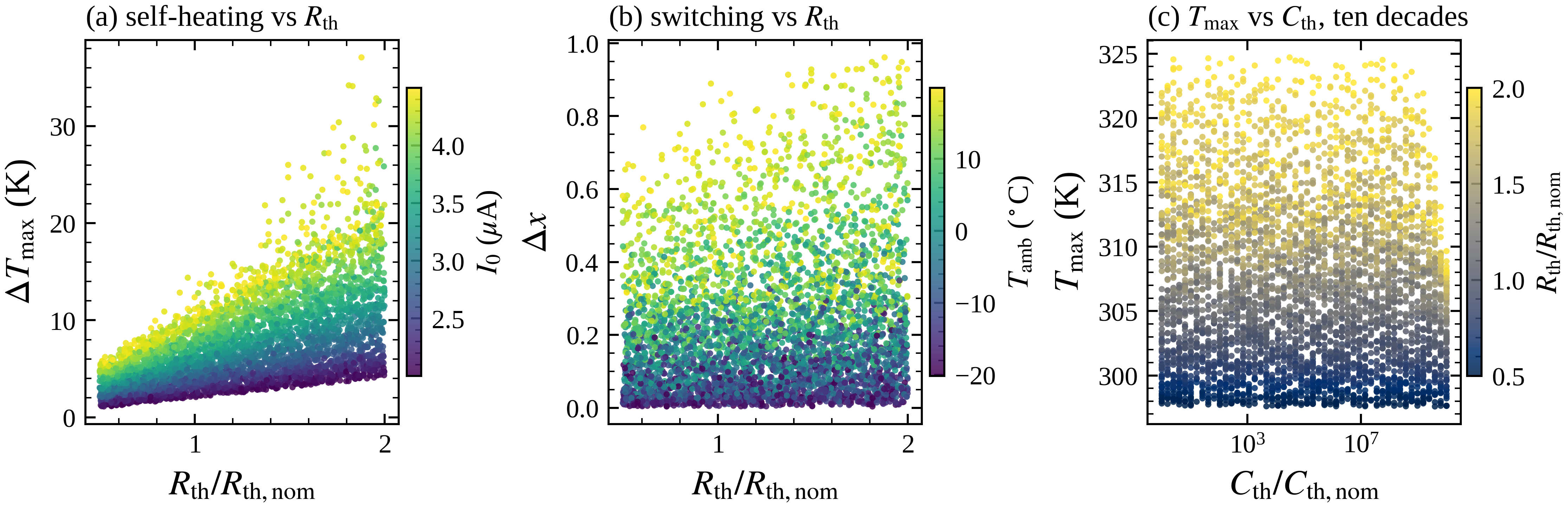}
\caption{Thermal-parameter response maps. (a) Self-heating against $\Rth$,
colored by $I_0$. (b) Excursion against $\Rth$, colored by $\Tamb$: the
bands are the ambient stratification and $\Rth$ scarcely tilts them.
(c) Peak temperature against $\Cth$ over ten decades, colored by $\Rth$, from
the full-factorial grid at fixed $I_0$ and $\Tamb$---stratified by $\Rth$ and
flat in $\Cth$.}
\label{fig:thermal-maps}
\end{figure*}

\Cref{fig:thermal-maps} isolates the two thermal parameters on the same kind
of axes and contrasts them directly. Self-heating rises linearly with $\Rth$,
at $6.3$~K per unit of $\Rth/R_{\rm th,nom}$ and stratified by drive
amplitude---the product structure \cref{eq:Rth-sensitivity} predicts, now over
the sampled range rather than at a single operating point. The same sweep
barely tilts the ambient bands of the excursion, which is
$S_{T_i}(\Rth,\Delta x)=0.042$ against $0.770$ for $\Tamb$ seen directly.
Panel (c) is the contrast that matters: over ten decades of $\Cth$ at fixed
drive and ambient the median peak temperature moves by $1.6$~K, against the
$27$~K spread that a factor four on $\Rth$ produces. \new{Over the ranges
studied, $\Rth$ therefore exerts a much stronger influence on self-heating
than $\Cth$}, and it is the visual counterpart of $\Cth$ being
absent from $\Theta$ in \cref{eq:Theta}.

\begin{figure}[!bt]
\centering
\includegraphics[width=9cm,height=9cm, trim = 0 0 0 0, clip]{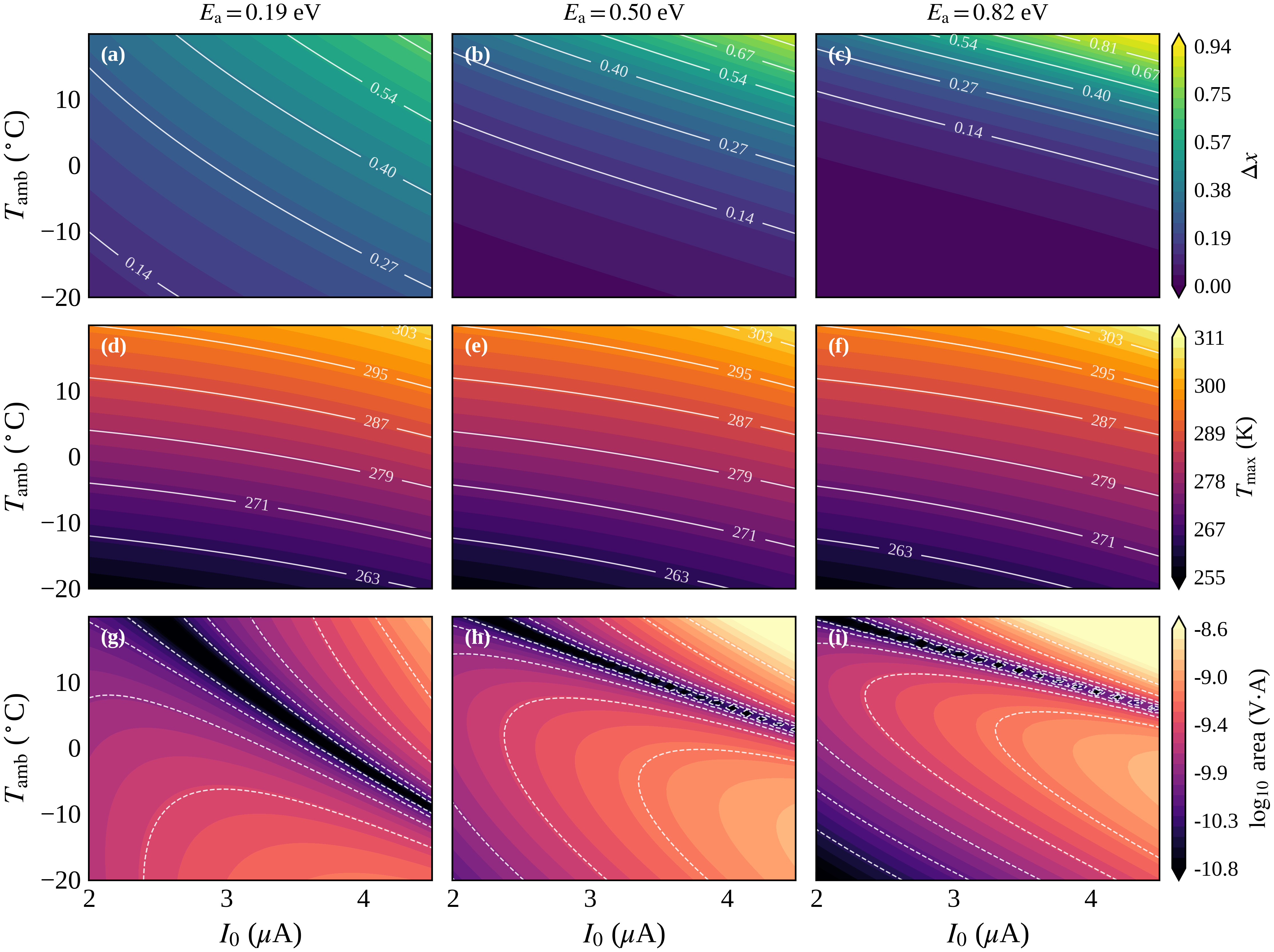}
\caption{Slices of the response surface in the $(I_0,\Tamb)$ plane at three activation energies from the regular $50^3$ grid. Rows show (a--c) excursion $\Delta x$, (d--f) peak temperature $T_{\max}$, and (g--i) loop area on a logarithmic scale. Each row shares one color scale. For (g--i), the scale is clipped to the $2$nd--$98$th percentile and crowded contours are omitted to highlight the valley migration.}
\label{fig:slices}
\end{figure}

The Saltelli sample is irregularly spaced and ill-suited to showing the
\emph{shape} of the response surface, so the three outputs were recomputed on
a regular full-factorial grid ($50$ values per parameter, $1.25\times10^{5}$
evaluations, same ranges and conditions). \Cref{fig:slices} presents that grid
as slices in the $(I_0,\Tamb)$ plane at three activation energies, one per
column, each row sharing a color scale so that the deformation with $\Ea$ is
read by moving across the row. 

The three rows say three different things. For
$\Delta x$ the contours shift and steepen markedly from column to column, and
run far closer to the $I_0$ axis than to the $\Tamb$ axis: ambient temperature
first, activation energy second. For $T_{\max}$ they are near-horizontal
\emph{and essentially identical across the three columns}---a mean absolute
difference of $1.5\,\%$ of the row's range between $\Ea=0.82$ and $0.19$~eV,
the geometric statement of $S_{T_i}(\Ea)\approx0.001$. The loop-area row
contains a narrow valley, the locus of near-degenerate loops already met in
\cref{sec:qs-reduction}, and that valley \emph{migrates} across the plane as
$\Ea$ grows, with a mean slice-to-slice difference of $25\,\%$ of the row's
range---seventeen times that of $T_{\max}$. This migrating degeneracy is what
the interaction-dominated indices of \cref{tab:sobol-4param} encode.

\section{Scope and limitations}
\label{sec:limitations}

\paragraph{Model scope and calibration.}
ATDM is a deliberately compact electrothermal extension of the Strukov drift model. Its lumped thermal node does not resolve spatial gradients, electrode and substrate heat spreading, or microscopic changes in the active region. Accordingly, \(R_{\mathrm{th}}\) and the representative activation energy \(E_a=0.7~\mathrm{eV}\) remain uncalibrated, and the predicted absolute temperatures are conditional on these choices. The present results establish the model's internal consistency, dimensionless structure, and parameter dependence; device-specific predictions require calibration against electrical and thermal measurements.

\paragraph{Operating regime and excitation.}
The baseline campaign lies in the current-driven quasi-static regime. The regime map quantifies the reduction error up to and beyond \(\varepsilon=1\), while the enlarged-area example of Appendix~C demonstrates access to the joint regime without claiming a calibrated device design. Because voltage drive changes the Joule power from \(i^2R(x)\) to \(v^2/R(x)\), the present scaling and feedback conclusions should not be transferred to voltage-driven operation without rederivation.

\paragraph{Statistical and physical scope.}
The loop area spans ten decades and is therefore analyzed through \(\log_{10}(\mathrm{area})\), which stabilizes the Sobol' estimates but does not fully separate the contributions of \(I_0\) and \(T_{\mathrm{amb}}\). A moment-independent sensitivity measure would provide a useful refinement. Microscopic defect interactions, stochastic fluctuations, structural degradation, and temperature-dependent thermal properties are outside the present compact-model scope.

\section{Conclusion}
\label{sec:conclusion}

ATDM extends the classical drift model of Strukov \textit{et al.} by coupling
its drift equation to a lumped heat balance through an Arrhenius-activated
ionic mobility, while retaining the original state variable and
series-resistor construction. The isothermal limit is recovered analytically
and, against an independently implemented reference, numerically \new{with
voltage discrepancies of order $10^{-4}\,\%$} of the peak voltage. This is limiting-case consistency, not
experimental validation: ATDM isolates the consequences of an assumed
electrothermal feedback within the Strukov picture.

The principal result is a scaling one. Recast in dimensionless form, the
switching response is \new{organized empirically, over the tested
quasi-static domain,} by two groups---a thermal lag $\varepsilon$
and an effective switching number $\Theta$ combining the isothermal drift per
half-period with the cycle-averaged Arrhenius enhancement, \new{the latter
through a closed-form approximation valid quantitatively for $\Lambda\lesssim1$}. Across $2.5\times10^{5}$ simulations spanning five physical
parameters, including the thermal capacitance over ten decades, the excursion
\new{is organized by $\Theta$ with a robust relative scatter of $4\,\%$} and no
fitted constant. \new{The weak influence of $\Cth$ in the tested
quasi-static regime is consistent with its absence
from $\Theta$ and is confirmed by the independent
capacitance sweeps.} Mapping the $(\varepsilon,\Theta)$ plane
converts the quasi-static validity boundary into a line of that plane---a
median $0.07\,\%$ error below $\varepsilon=10^{-2}$ against $560\,\%$ beyond
$\varepsilon=1$---and bounds the region \new{reachable under the chosen drive,
geometry, and self-heating constraints}.

Two further results follow from the same formulation. Dynamically, \new{under current drive, the
stroboscopic map of the quasi-statically reduced model is strictly increasing,
which excludes period-doubling and chaos structurally for that reduction}; its measured Floquet multiplier stays
inside the unit circle over the ranges tested, \new{so the period-1 orbit is
locally attracting there, while its uniqueness is a separate numerical
finding}, and the multiplier fixes the \new{asymptotic} settling time a
stroboscopic study must respect---a physical time of order $300$~s at $I_0=5\,\mu$A, not a
period count. \new{For the same reduced model, this structure also rules out a sustained
ratchet drift of the kind that filamentary models with exponentially gated
kinetics can exhibit.} Statistically, the
variance-based analysis over $(I_0,\Tamb,\Ea,\Rth)$ shows the loop area to be
interaction-dominated, self-heating to be governed jointly by $I_0$ and
$\Rth$, and, on a properly conditioned metric, the activation energy to
outrank the thermal resistance for the loop---a ranking the raw metric
reverses.

\new{The model has also been implemented in executable form. A \textsc{verilog-a}
compact model, compiled with OpenVAF and run as a device inside
\texttt{ngspice}, reproduces the reference implementation to within
$0.009\,\%$ of the peak voltage across four amplitudes and both activation
energies, and returns the classical isothermal excursion at $\Ea=0$
(\cref{app:veriloga}). The isothermal limit is therefore established
analytically, and numerically under two independent integration engines.}

What remains unestablished is quantitative agreement with a specific device.
Both $\Ea$ and, more critically, $\Rth$ are uncalibrated. Experimental
calibration or defensible bounds for the two, and comparison against measured
current--voltage, switching-time, and thermal data, are the natural next step,
building on the verified reduction, the quantified sensitivity, and the
\new{simulator-ready compact model developed here}.

\section*{Funding Declaration}
N.G.K. appreciates financial support from the FAPESP--UNESCO-TWAS Project
(Grant No. 2024/08346-8).

\section*{Acknowledgments}
HAC thanks FAPESP grant 2021/14335-0 of the ICTP--SAIFR for partial support.

\section*{Declarations}

\subsection*{Conflict of interest / Competing interests}
The authors declare that they have no conflict of interest and no competing
interests.

\subsection*{Ethics approval and consent to participate}
Not applicable.

\subsection*{Data availability}
No experimental data were used in this study. All results were obtained by
numerical simulation of the model described in this article. The simulation
data, the Python implementation of the model and the scripts used to produce
the figures and tables, as well as the \textsc{Verilog-A} compact model and
the \texttt{ngspice} test benches, are available from the corresponding
author upon reasonable request.

\section*{Author Contributions}
N.G.K. conceived the study and developed the ATDM formulation, developed the numerical implementation, performed the
electrothermal simulations, carried out the global sensitivity
analysis, and implemented the \new{\texttt{Verilog-A/OSDI} compact model}.
N.G.K. performed the data analysis and visualization and drafted the manuscript. All authors contributed to the interpretation of the results, critically reviewed the manuscript, and approved the final version.

\section*{Appendix}
\begin{appendices}
\crefalias{section}{appendix}
\crefalias{subsection}{subappendix}

\section{Governing equations and nomenclature}
\label{app:equations}

\begin{equation}
\begin{aligned}
x(t)&=w(t)/D, \qquad R(x)=\Ron x+\Roff(1-x),\\
\muv(T)&=\mu_0\exp\!\Big[-\tfrac{\Ea}{\kb}\big(\tfrac1T-\tfrac1{T_0}\big)\Big],\\
\dot x &= \tfrac{\muv(T)\Ron}{D^2}\,i\,F(x,i), \quad
\Cth\dot T = i^2R(x)-\tfrac{T-\Tamb}{\Rth}.
\end{aligned}
\end{equation}

\section{The isothermal limit depends on $I_0/f$ alone}
\label{app:scaling}

\begin{figure}[ht!]
\centering
\includegraphics[width=8.3cm,height=3.6cm, trim = 0 0 0 0, clip]{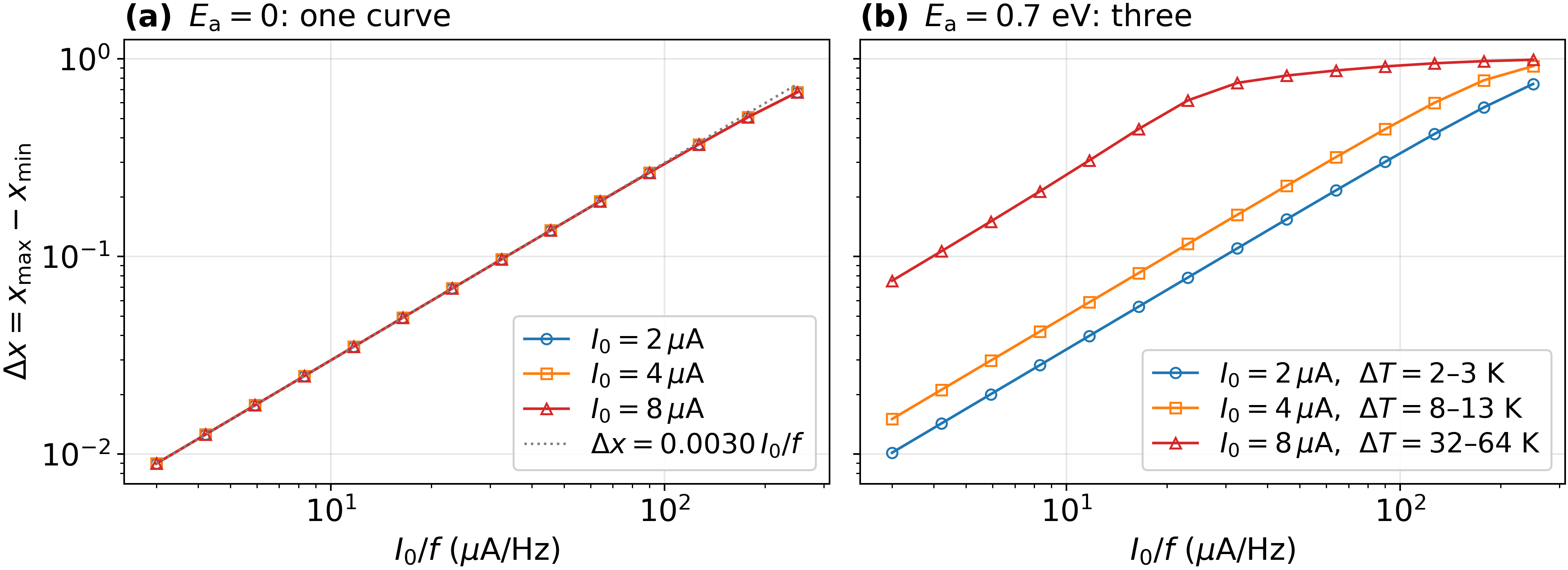}
\caption{Excursion $\Delta x$ of the settled orbit against $I_0/f$ for three
amplitudes. (a) Isothermal limit, $\Ea=0$: the three currents fall on a single
curve. (b) At $\Ea=0.7$~eV the curves separate in order of current, the peak
temperature rise being given in the legend. Every point is a settled orbit,
the residual drift per cycle staying below $10^{-4}$ of $\Delta x$.}
\label{fig:scaling}
\end{figure}

\Cref{fig:scaling} tests \cref{eq:Pi} on the raw orbit rather than on an
aggregate metric. At constant mobility the drift depends on $I_0$ and $f$ only
through their ratio, and panel (a) confirms it: three amplitudes spanning a
factor of four fall on one curve, $\Delta x=0.00298\,I_0/f$, with a relative
spread of at most $8\times10^{-6}$ over the fourteen nonsaturated points.

The slope corresponds to a mean window value $\langle F\rangle=0.936$ over the
half-cycle, so the Biolek window slows the drift by only six percent at these
operating points. Panel (b) repeats the sweep at $\Ea=0.7$~eV: the single
curve splits into three, ordered by current, because the Joule power scales as
$I_0^2$ and the resulting temperature rise---$2$--$3$~K at $2\,\mu$A against
$32$--$64$~K at $8\,\mu$A---does not collapse onto $I_0/f$. At equal $I_0/f$
the excursion then differs by a factor $1.3$ to $7.9$ between the two. This is
the direct demonstration that a second group is required, and it is the
orbit-level counterpart of \cref{fig:collapse}.

\begin{table}[!ht]
\centering
\footnotesize
\caption{Symbols used throughout.}
\label{tab:nomenclature}
\begin{tabular}{lll}
\toprule
Symbol & Meaning & Unit\\
\midrule
$x$ & Normalized state variable & --\\
$D$ & Oxide thickness & nm\\
$\Ron$, $\Roff$ & Low-/high-resistance states & $\Omega$\\
$i$, $v$ & Device current, voltage & A, V\\
$T$, $\Tamb$, $T_0$ & Device, ambient, reference temp. & K\\
$\Rth$, $\Cth$ & Thermal resistance, capacitance & K/W, J/K\\
$\muv$ & Ionic mobility & m$^2$V$^{-1}$s$^{-1}$\\
$\Ea$ & Activation energy & eV\\
$\kb$ & Boltzmann constant & eV/K\\
$F(x,i)$ & Biolek window function & --\\
$p$ & Window sharpness parameter & --\\
$\tau_{\rm th}$ & Thermal time const. ($=\Rth\Cth$) & s\\
$\varepsilon$ & Thermal lag $\tau_{\rm th}/\tau_P$ & --\\
$\Pi$, $\Theta$ & Switch numb., \cref{eq:Pi,eq:Theta} & --\\
$\lambda$ & Floquet multiplier & --\\
\bottomrule
\end{tabular}
\end{table}

\section{Beyond the quasi-static boundary}
\label{app:beyond-qs}

\begin{figure}[ht!]
\centering
\includegraphics[width=8.3cm,height=3.6cm, trim = 0 0 0 0, clip]{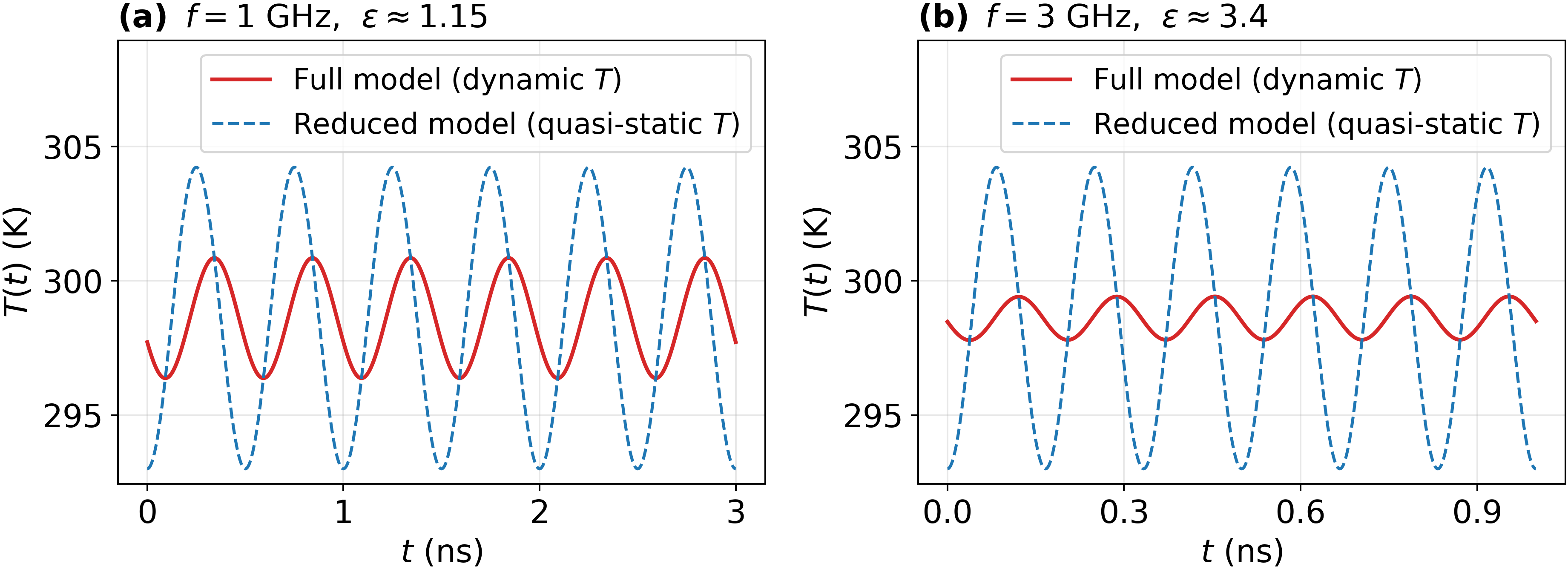}
\caption{Temperature predicted by the full model (red, dynamic $T$) and the
reduced one (blue dashed, algebraic $T$) at $I_0=4\,\mu$A, $\Ea=0.7$~eV, run
independently at $f=1$~GHz (a), $\varepsilon\approx1.15$, and $3$~GHz (b),
$\varepsilon\approx3.4$. The reduced amplitude is frequency-independent by
construction; the full response is damped and phase-lagged, collapsing to
$4.5$ and $1.6$~K.}
\label{fig:full-vs-reduced-ghz}
\end{figure}

\paragraph{Temperature at GHz frequencies.} \Cref{fig:full-vs-reduced-ghz}
runs the full and reduced formulations independently at $I_0=4\,\mu$A beyond
the $\varepsilon=1$ threshold. At $f=1$~GHz ($\varepsilon\approx1.15$) the
reduced model overshoots the temperature swing by a factor $\sim2.5$
($11.2$~K predicted against $4.5$~K actual) and lags in phase. At $3$~GHz
($\varepsilon\approx3.4$) its prediction is unchanged in amplitude---the
reduced equation has no memory of the forcing timescale---while the full
model's response has collapsed to $1.6$~K, an error of a factor $\sim7$.
Beyond the threshold the reduction does not merely lose accuracy; it becomes
qualitatively wrong.

\begin{figure}[ht!]
\centering
\includegraphics[width=8.3cm,height=3.8cm, trim = 0 0 0 0, clip]{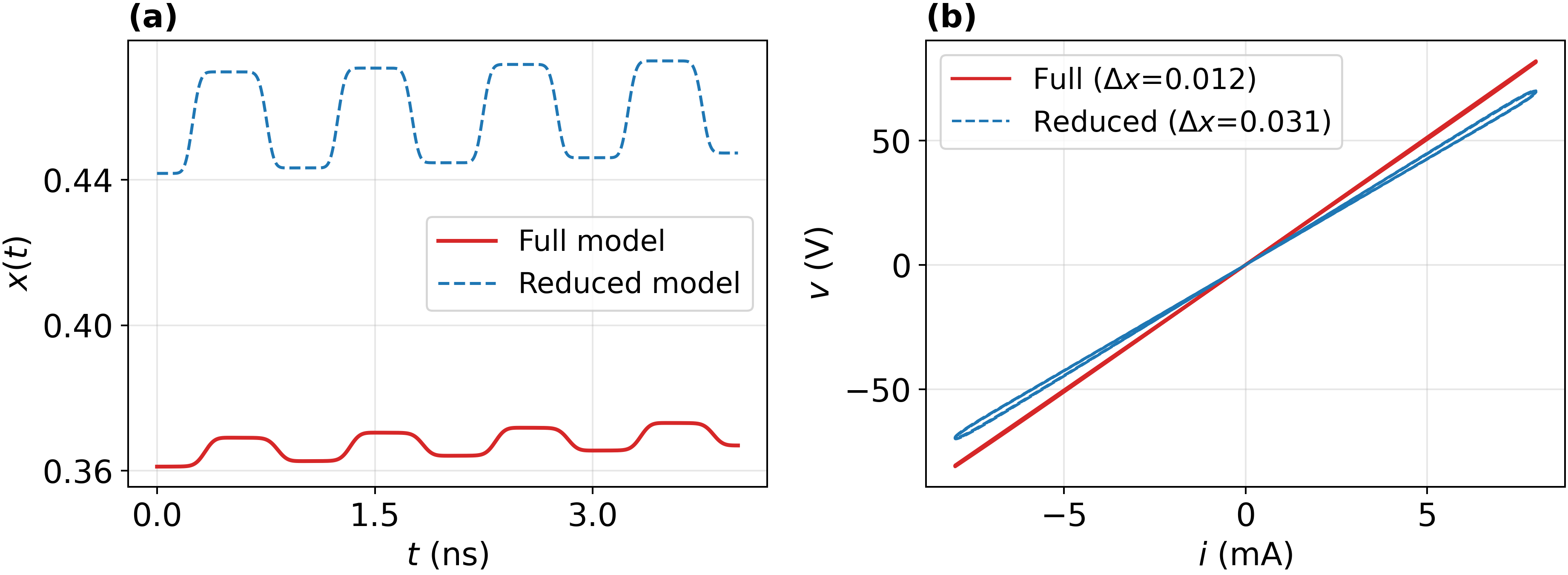}
\caption{The joint regime, reached by enlarging the active area to
$10^5\times$ baseline at $I_0=8$~mA, $f=1$~GHz, $\Ea=0.7$~eV. (a) State
variable for the full (red) and reduced (blue dashed) models, run
independently: different operating points and an excursion overestimated by
$2.6$--$3.2\times$. (b) The resulting $i$--$v$ loops, nearly linear at this
excursion, differing mainly in slope.}
\label{fig:joint-regime}
\end{figure}

\paragraph{A regime where the second state changes switching.} At those
frequencies and the baseline area the excursion itself is negligible
($\Delta x\lesssim10^{-9}$), so the dynamic state matters for $T(t)$ while
$x(t)$ is too frozen for the improved temperature to feed back into anything.
Since $\Rth=D/(\kappa A)$ while $\tau_{\rm th}$ is area-independent, enlarging
$A$ lowers the temperature rise at a given current without shifting the
$\varepsilon=1$ threshold: a larger device can carry more current, and switch
faster, at the same bounded temperature. Scanning $A$ and $I_0$ jointly at
$1$~GHz locates such a point at $A=10^5\times$ baseline
($3.16\,\mu$m square, a crossbar-scale device) and $I_0=8$~mA
(\cref{fig:joint-regime}): the full model settles into a small-amplitude
oscillation around $x\approx0.36$, the reduced model predicts a different
operating point ($x\approx0.45$) and overestimates $\Delta x$ by
$2.6$--$3.2\times$. The predicted temperature there ($580$--$640$~K) is
elevated, so this is a feasibility argument rather than a device-specific
prediction.

\begin{figure}[ht!]
\centering
\includegraphics[width=8.5cm,height=4cm, trim = 0 0 0 0, clip]{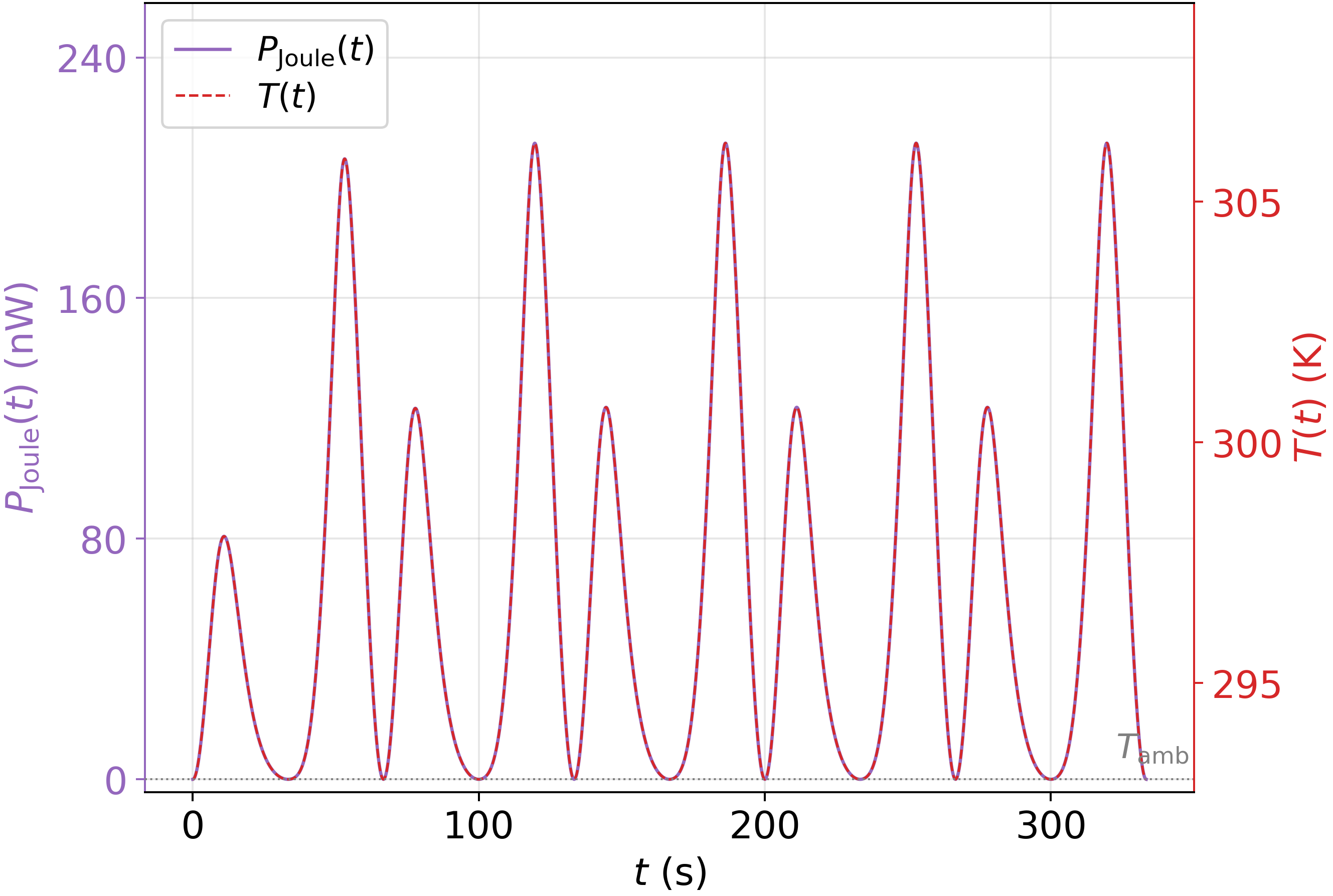}
\caption{\new{Instantaneous Joule power $P_{\rm Joule}(t)$ (solid, left axis)
and device temperature $T(t)$ (dashed, right axis) over five forcing periods
at the baseline operating point ($I_0=4\,\mu$A, $f=0.015$~Hz,
$\Ea=0.7$~eV). The two axes are related by $T=\Tamb+\Rth P_{\rm Joule}$ and
the curves coincide at plotting resolution. After the first forcing period,
which still carries the initial transient (small peak at $\sim80$~nW and a
slightly lower large peak), the settled response alternates between
peaks of $\sim120$ and $\sim211$~nW in each period.}}
\label{fig:joule-T}
\end{figure}

\paragraph{Instantaneous Joule power.} \new{Read at the level of
$P_{\rm Joule}(t)=i^2(t)R(x(t))$ rather than of the temperature it drives, the
mechanism of \cref{sec:joule-3a} is an alternation between $\sim120$ and
$\sim211$~nW peaks per settled forcing period (\cref{fig:joule-T}), which is the frequency doubling of
\cref{fig:xT} peak for peak: in the quasi-static regime the two are tied by
$T=\Tamb+\Rth P_{\rm Joule}$, so the temperature trace carries no information
the power trace does not. The $\sim80$~nW peak visible at the start of
\cref{fig:joule-T} belongs to the first half-cycle, before the state has
reached its periodic orbit, and is not representative of the settled
response.}

\section{Sensitivity analysis: method and supporting results}
\label{app:sobol}

\paragraph{Estimators and sampling.} For $D$ inputs and a scalar output $Y$,
the first-order index $S_i$ is the fraction of $\mathrm{Var}(Y)$ explained by
parameter $i$ alone and the total-order index $S_{T_i}$ the fraction explained
by parameter $i$ together with all its interactions, so that $S_{T_i}-S_i$
isolates coupling. Following Saltelli \cite{Saltelli2010}, two independent
Sobol' sequences $A,B\in[0,1]^{N\times D}$ are drawn with $N=1024=2^{10}$---
a power of two, so the sequences keep their equidistribution---and scaled to
$I_0\in[2,4.5]\,\mu$A, $\Tamb\in[-20,20]\,^\circ$C, $\Ea\in[0.19,0.82]$~eV and
$\Rth\in[0.5,2]\times$ baseline. Matrices $AB_i$ replace column $i$ of $A$ by
that of $B$, and
\begin{equation}
S_i=\frac{\frac{1}{N}\sum_j y_B(j)\big(y_{AB_i}(j)-y_A(j)\big)}{\mathrm{Var}(y)},
\end{equation}
\begin{equation}
S_{T_i}=\frac{\frac{1}{2N}\sum_j \big(y_A(j)-y_{AB_i}(j)\big)^2}{\mathrm{Var}(y)},
\label{eq:sobol-estimators}
\end{equation}
with $\mathrm{Var}(y)$ from the pooled sample. All values reported here use
the centered estimator: the first-order estimator is invariant under
$y\mapsto y+c$ only in expectation, and in a finite sample the shift
contributes a term of order $c/\sqrt N$ which, for an output such as
$T_{\max}$ whose mean is twenty times its standard deviation, dominates
completely---without centring the bootstrap interval for $S_i(\Tamb)$ spans
$[-0.97,+2.67]$, against $[+0.80,+0.94]$ once the pooled mean is subtracted.

\paragraph{Integration engine.} The $6144$ evaluations use a vectorized
fixed-step semi-implicit scheme ($\mathrm{d}t=0.02$~s, $30$ periods) rather
than the adaptive Radau integration used elsewhere, which makes a sample of
this size tractable. \Cref{tab:validation} verifies it against the Radau
reference at thirteen control points spanning the ranges: the error on
$\Delta x$ stays at or below $0.024\,\%$, on $T_{\max}$ below $0.011$~K, and
on the loop area below $0.043\,\%$ of the grid's full-scale maximum. The
relative area error reaches double digits only where the area itself is five
orders of magnitude below full scale---the same ill-conditioning discussed in
\cref{sec:sobol}.

\paragraph{Three-parameter ensemble.} \Cref{tab:sobol} reports the same
analysis at fixed baseline $\Rth$, over $(I_0,\Tamb,\Ea)$ only. The rankings
of \cref{tab:sobol-4param} are already present there: adding $\Rth$ does not
overturn the picture for $\Delta x$, $T_{\max}$, or the $I_0$/$\Tamb$-driven
part of the loop area, but it does take a large share of $\Delta T_{\max}$,
which the three-parameter analysis attributes almost entirely to $I_0$ by
construction.

\begin{table}[!h]
\centering
\scriptsize
\caption{Fixed-step engine against the Radau reference at thirteen control
points: nine low/median/high sweeps of each parameter with the others at their
midpoint, and four points in the low-area corner. Area error is normalized by
the grid's full-scale maximum $A_{\max}=1.857\times10^{-8}$~V$\cdot$A.}
\label{tab:validation}
\begin{tabular}{lccccc}
\toprule
Point & $I_0$ & $\Tamb$ & $\Ea$ & $\Delta x$ err. & area err.\\
 & ($\mu$A) & ($^\circ$C) & (eV) & (\%) & (\% of $A_{\max}$)\\
\midrule
$I_0$ low      & 2.00 & 0.0    & 0.505 & 0.003 & 0.00014\\
$I_0$ median   & 3.25 & 0.0    & 0.505 & 0.000 & 0.00334\\
$I_0$ high     & 4.50 & 0.0    & 0.505 & 0.024 & 0.03862\\
$\Tamb$ low    & 3.25 & $-20.0$& 0.505 & 0.002 & 0.00021\\
$\Tamb$ high   & 3.25 & 20.0   & 0.505 & 0.014 & 0.04293\\
$\Ea$ low      & 3.25 & 0.0    & 0.190 & 0.008 & 0.00497\\
$\Ea$ high     & 3.25 & 0.0    & 0.820 & 0.003 & 0.00154\\
\midrule
Low-area 1 & 2.00 & $-20.0$ & 0.190 & 0.003 & 0.00010\\
Low-area 2 & 2.00 & 20.0    & 0.190 & 0.009 & 0.00060\\
Low-area 3 & 2.25 & $-20.0$ & 0.300 & 0.003 & 0.00014\\
Low-area 4 & 2.25 & 0.0     & 0.250 & 0.005 & 0.00049\\
\bottomrule
\end{tabular}
\end{table}

\begin{table}[!h]
\centering
\footnotesize
\caption{Sobol' indices over $(I_0,\Tamb,\Ea)$ at fixed baseline $\Rth$;
$N=1024$, $5120$ evaluations.}
\label{tab:sobol}
\begin{tabular}{llccc}
\toprule
Output & Parameter & $S_i$ & $S_{T_i}$ & $S_{T_i}-S_i$\\
\midrule
\multirow{3}{*}{$\Delta x$}
 & $I_0$   & $0.142$ & $0.211$ & $0.069$ \\
 & $\Tamb$ & $0.703$ & $0.798$ & $0.095$ \\
 & $\Ea$   & $0.060$ & $0.088$ & $0.028$ \\
\midrule
\multirow{3}{*}{$\Delta T_{\max}$}
 & $I_0$   & $0.968$ & $0.984$ & $0.016$ \\
 & $\Tamb$ & $0.012$ & $0.032$ & $0.019$ \\
 & $\Ea$   & $0.003$ & $0.010$ & $0.007$ \\
\midrule
\multirow{3}{*}{Loop area}
 & $I_0$   & $0.151$ & $0.760$ & $0.609$ \\
 & $\Tamb$ & $0.164$ & $0.752$ & $0.588$ \\
 & $\Ea$   & $0.022$ & $0.369$ & $0.346$ \\
\bottomrule
\end{tabular}
\end{table}

\begin{figure*}[!b]
\centering
\includegraphics[width=16cm,height=4cm,trim=0 0 0 0,clip]{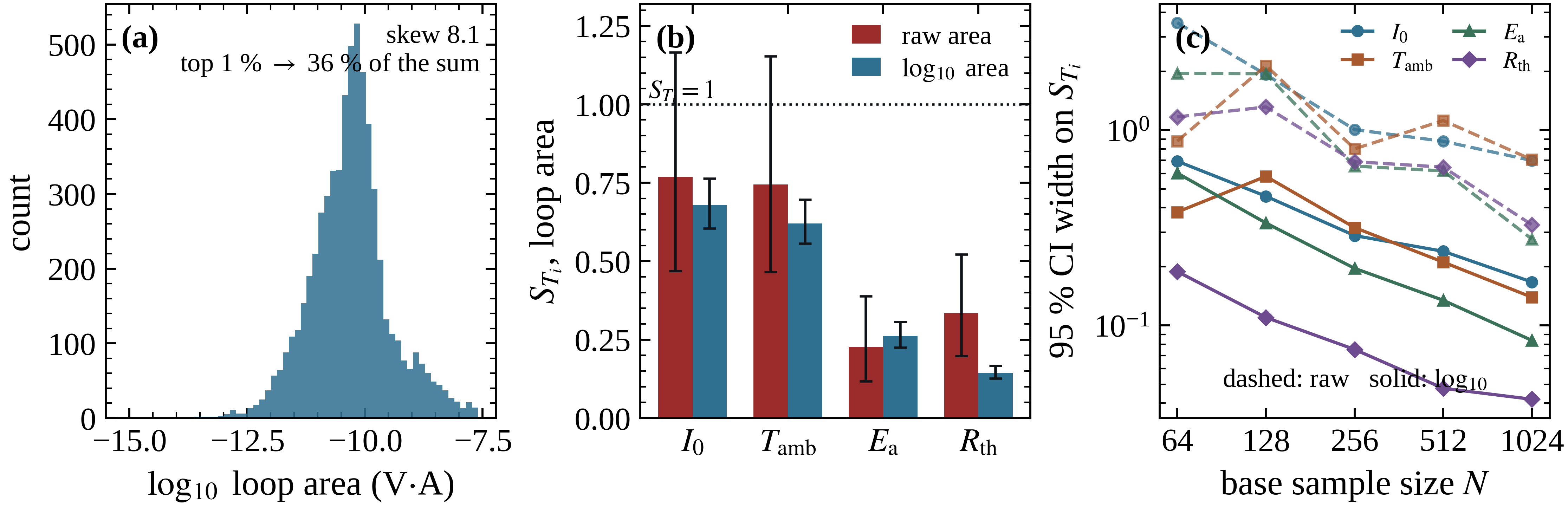}
\caption{Conditioning of the loop-area sensitivity analysis. (a) Distribution
of the loop area on a logarithmic abscissa: the raw quantity has skewness
$8.1$ and its largest $1\,\%$ of points carry $36\,\%$ of the total, while its
logarithm is near-symmetric. (b) Total-order indices with $95\,\%$ bootstrap
intervals, raw against $\log_{10}$. (c) Interval width against $N$ for both
metrics; dashed, raw; solid, $\log_{10}$.}
\label{fig:sobol-logarea}
\end{figure*}

\paragraph{Uncertainty and conditioning.} A percentile bootstrap with $2000$
resamples of the $N=1024$ base samples, reusing the stored outputs and
requiring no further evaluation, gives the intervals quoted in
\cref{sec:sobol}; the same resampled index set is applied to $y_A$, $y_B$, and
every $y_{AB_i}$, so each replicate remains a valid Saltelli ensemble. With
intervals attached, the negative first-order estimates of the tables resolve
as estimation noise, $S_i(\Ea,T_{\max})=-0.000\,[-0.003,+0.003]$. Repeating
the estimation for $N\in\{64,\dots,1024\}$ by truncating each Saltelli block
---not by random subsampling, which would destroy the equidistribution---
narrows the interval on $S_{T_i}(I_0)$ for the loop area from $3.55$ to
$0.70$, with the ranking unchanged from $N=256$ onward. \Cref{tab:logarea} and
\cref{fig:sobol-logarea} then compare the raw and $\log_{10}$ metrics: the
interval widths narrow by factors of three to eight and every upper bound
returns below unity. By the $1/\sqrt N$ scaling, the factor $4.4$ gained on
$S_{T_i}(I_0)$ is worth roughly $19\times$ the sample size, so $\log_{10}$ at
$N=64$ is already as tight as the raw metric at $N=1024$.

\begin{table}[!h]
\centering
\footnotesize
\caption{Total-order indices for the hysteresis loop area, raw against
$\log_{10}$-transformed, with $95\,\%$ bootstrap intervals ($N=1024$, $D=4$,
centered estimator).}
\label{tab:logarea}
\begin{tabular}{lcccc}
\toprule
& \multicolumn{2}{c}{raw area} & \multicolumn{2}{c}{$\log_{10}$ area}\\
\cmidrule(lr){2-3}\cmidrule(lr){4-5}
Parameter & $S_{T_i}$ & $95\,\%$ CI & $S_{T_i}$ & $95\,\%$ CI\\
\midrule
$I_0$   & $0.768$ & $[0.469,1.166]$ & $0.679$ & $[0.604,0.763]$\\
$\Tamb$ & $0.745$ & $[0.466,1.154]$ & $0.621$ & $[0.557,0.696]$\\
$\Ea$   & $0.226$ & $[0.118,0.389]$ & $0.262$ & $[0.225,0.307]$\\
$\Rth$  & $0.334$ & $[0.198,0.522]$ & $0.145$ & $[0.127,0.167]$\\
\bottomrule
\end{tabular}
\end{table}

\section{Verification of the fast integration engine}
\label{app:engine}

The regime map of \cref{sec:regimemap} requires both formulations to be
integrated at several thousand operating points, which the adaptive Radau
solver makes impractical. A fixed-step engine is used instead and verified
here against that reference rather than asserted equivalent.

The stiffness of \cref{eq:system3a} is entirely the linear thermal relaxation
at rate $1/\tau_{\rm th}\approx5.5\times10^{9}\,\mathrm{s}^{-1}$. That term is
integrated \emph{exactly} over each step, holding the Joule power constant
across it,
\begin{equation}
T_{n+1}=\Tamb+\Rth P_n+\big(T_n-\Tamb-\Rth P_n\big)\,
e^{-\Delta t/\tau_{\rm th}},
\label{eq:expstep}
\end{equation}
which is unconditionally stable for any $\Delta t$, reduces to the
quasi-static answer when $\Delta t\gg\tau_{\rm th}$ and resolves the thermal
lag when $\Delta t\lesssim\tau_{\rm th}$. The state variable uses an explicit
trapezoidal step on top of it, so no implicit solve is required and the cost
per operating point becomes independent of frequency. Both formulations use
the same stepper, so the comparison in \cref{fig:regimemap} is not
contaminated by a difference of solver.

\begin{figure*}[ht!]
\centering
\includegraphics[width=16cm,height=8cm, trim = 0 0 0 0, clip]{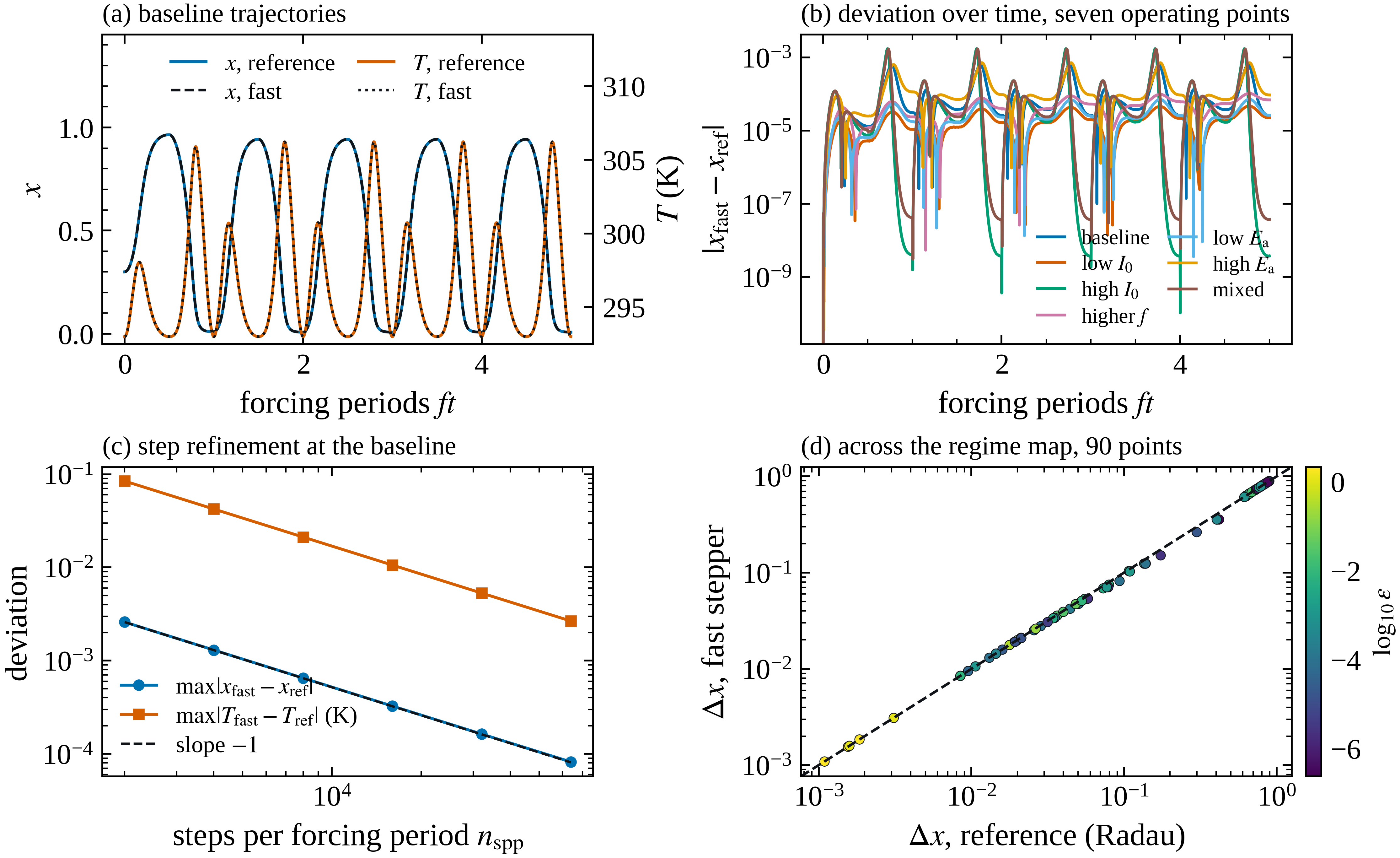}
\caption{Fixed-step engine against the adaptive Radau reference.
(a) Trajectory overlay at the baseline operating point. (b) Deviation over
time at seven operating points. (c) Step refinement: the deviation halves with
each halving of the step, the signature of discretization error rather than a
difference of model. (d) The same comparison across $90$ points of the regime
map, colored by $\varepsilon$; the dashed line is $1{:}1$.}
\label{fig:engine}
\end{figure*}

\Cref{fig:engine} reports the verification. Over seven operating points
spanning the ranges used here, $\Delta x$ agrees with the reference to four
decimal places and $\Delta T_{\max}$ to $0.01$~K on rises of $13$ to $35$~K,
with no drift over five forcing periods. Refining the step gives the decisive
test: the deviation falls by exactly a factor of two for each halving of
$\Delta t$, converging to $\Delta x=0.93526$ and $\Delta T_{\max}=13.2194$~K.
A residual that decreases with $\Delta t$ establishes that the two engines
integrate the same equations and that the gap is discretization alone; a
constant residual would have indicated otherwise. Carried across the regime
map, with $90$ points reintegrated using the reference engine, the median
relative deviation on $\Delta x$ is $0.22\,\%$.

\section{Robustness of the reduction criteria and of the window choice}
\label{app:robustness}

Three quantities the main text relies on are checked here against the
assumptions that produced them: the closed form of the cycle-averaged
Arrhenius factor, the measured power timescale $\tau_P$, and the window
function.

\paragraph{The closed-form $\langle\Gamma\rangle$.} \Cref{eq:Gamma} assumes a
$\sin^{2}$ temperature waveform. \Cref{tab:gamma-check} compares it with the
exact cycle average of $\muv(T(t))/\mu_0$, computed by trapezoidal quadrature
on the last period of a settled orbit ($60$ forcing periods, $600$ samples per
period). \new{For $\Lambda\lesssim1$ the median absolute relative error of the closed
form is $1.2\,\%$, reaching $5.2\,\%$ at the points tested}, and it underestimates the enhancement increasingly beyond,
because the true temperature waveform is fuller than $\sin^{2}$ and because
$R(x(t))$ modulates the power within the cycle. The domain of quantitative
validity of $\Theta$ is therefore $\Lambda\lesssim1$; outside it $\Theta$
remains a monotone ordering variable, which is what the collapse uses.

\begin{table}[ht!]
\centering
\footnotesize
\caption{Closed-form $G_{\rm amb}\langle\Gamma\rangle$ against the exact cycle
average of $\muv(T(t))/\mu_0$ on settled orbits.}
\label{tab:gamma-check}
\begin{tabular}{cccccc}
\toprule
$I_0$ & $f$ & $\Tamb$ & $\Lambda$ & closed & error\\
($\mu$A) & (Hz) & (K) & & / exact & (\%)\\
\midrule
$2$ & $0.015$ & $293$ & $0.19$ & $1.102/1.103$ & $-0.1$\\
$4$ & $0.015$ & $293$ & $0.74$ & $1.499/1.561$ & $-4.0$\\
$4$ & $0.100$ & $293$ & $0.74$ & $1.499/1.507$ & $-0.5$\\
$4$ & $0.500$ & $293$ & $0.74$ & $1.499/1.518$ & $-1.2$\\
$4$ & $0.015$ & $253$ & $0.99$ & $0.0218/0.0230$ & $-5.2$\\
$5$ & $0.015$ & $293$ & $1.14$ & $1.916/2.290$ & $-16.3$\\
$8$ & $0.100$ & $293$ & $2.75$ & $6.03/12.54$ & $-51.9$\\
\bottomrule
\end{tabular}
\end{table}

\paragraph{The power timescale $\tau_P$.} \Cref{tab:tauP-check} measures
$\tau_P$ at fixed $f=0.015$~Hz for three current waveforms and two solver
tolerances. \new{Tightening the tolerance by three decades changes $\tau_P$ by about
$0.01\,\%$ for the sine drive and by about $1.5\,\%$ and $2.3\,\%$ for the
triangular and smoothed square drives, while $\tau_P f$ changes by at most one
unit in its last reported digit; these variations are small compared with the
order-of-magnitude change produced by the waveform, so the criterion is not an
artifact of the integrator.} Changing
the waveform moves it by an order of magnitude at unchanged frequency, which
is the point of measuring it rather than substituting $1/f$: the thermal node
must follow the power, and a square-ish drive makes the power vary far faster
than the period.

\begin{table}[ht!]
\centering
\footnotesize
\caption{Measured power timescale and thermal lag at $I_0=4\,\mu$A,
$f=0.015$~Hz, $\Ea=0.7$~eV, for three waveforms and two solver tolerances.}
\label{tab:tauP-check}
\begin{tabular}{lcccc}
\toprule
Waveform & rtol & $\tau_P$ (s) & $\tau_Pf$ & $\varepsilon$\\
\midrule
sine & $10^{-6}$ & $8.8720$ & $0.133$ & $2.06\times10^{-11}$\\
sine & $10^{-9}$ & $8.8709$ & $0.133$ & $2.06\times10^{-11}$\\
triangle & $10^{-6}$ & $3.9528$ & $0.059$ & $4.63\times10^{-11}$\\
triangle & $10^{-9}$ & $3.8946$ & $0.058$ & $4.70\times10^{-11}$\\
square (smoothed) & $10^{-6}$ & $0.7128$ & $0.011$ & $2.57\times10^{-10}$\\
square (smoothed) & $10^{-9}$ & $0.6962$ & $0.010$ & $2.63\times10^{-10}$\\
\bottomrule
\end{tabular}
\end{table}

\paragraph{The window function.} Several statements in the main text could in
principle be artifacts of the Biolek window. Replacing it by the Joglekar
window $F=1-(2x-1)^{2p}$, which is symmetric and independent of the sign of
the current, separates what survives from what does not
(\cref{tab:window-check}).

\begin{table}[!b]
\centering
\footnotesize
\caption{Biolek against Joglekar window: what changes. Poincar\'e quantities
at $I_0=5\,\mu$A, $f=0.10$~Hz; excursion symmetry at $I_0=4\,\mu$A from
$x(0)=0.3$; isothermal collapse at $I_0/f=100\,\mu$A/Hz.}
\label{tab:window-check}
\begin{tabular}{lcc}
\toprule
Quantity & Biolek & Joglekar\\
\midrule
$P$ increasing & yes & yes\\
$\mathrm{d}P/\mathrm{d}x$ & $0.385$--$0.753$ & $1.000$ (identity)\\
Fixed points & one, $\lambda=0.658$ & neutral continuum\\
$\langle x\rangle$, $f=0.02$--$1$~Hz & $0.500$ & $0.695$--$0.313$\\
$\Delta x\,f$ (linear regime) & $0.020$ & $0.013$--$0.026$\\
$\Delta x$ at $I_0/f=100$, $\Ea=0$ & $0.29420$ & $0.31720$\\
\quad (identical across $I_0$, $f$) & yes & yes\\
\bottomrule
\end{tabular}
\end{table}

The isothermal collapse on $I_0/f$ survives
unchanged---three pairs sharing $I_0/f=100\,\mu$A/Hz give the same excursion
to five decimal places with either window, only the prefactor differing
($0.29420$ for Biolek against $0.31720$ for Joglekar)---because the collapse
is a property of the drift equation and the prefactor a property of the mean
window value. Monotonicity of the stroboscopic map also survives, as
\cref{prop:monotone} requires. But the strict contraction does not: with the
Joglekar window the period map is the identity to within $10^{-6}$, every
state is a neutral fixed point, and consequently the settled orbit retains the
memory of its initial condition---$\langle x\rangle$ takes the values
$0.695$, $0.515$, $0.362$ and $0.313$ at $f=0.02$, $0.05$, $0.2$ and $1$~Hz
from $x(0)=0.3$, instead of the $0.500$ that the Biolek window enforces---and
the $1/f$ law for the excursion loses its constant prefactor
($\Delta x\,f=0.013$ to $0.026$ instead of $0.020$). Attractor selection in
ATDM is thus a property of the sign-switching window; the exclusion of
period-doubling and chaos is a property of the scalar reduction. Both
statements are made separately in \cref{sec:poincare} for that reason.

\section{\new{Verilog-A compact model}}
\label{app:veriloga}
\new{ATDM is supplied as the \textsc{Verilog-A} compact model
\texttt{atdm.va}, usable as a two-terminal circuit device.
It directly transcribes the window \cref{eq:biolek}, Arrhenius
factor \cref{eq:arrhenius}, resistance law \cref{eq:Rmem},
and relation $v=iR(x)$ from \cref{eq:system3a}; implementation
choices concern only the representation of the two states.}

\new{Internal nodes carry $x$ and $T-\Tamb$, with the drift
equation at the state node and the heat balance implemented
through the physical $\Rth$--$\Cth$ pair. The simulator's
local-truncation-error control handles the thermal stiffness
without adjustment of the integration options:
$\tau_{\rm th}=1.8\times10^{-10}$~s is about eleven orders
of magnitude below the forcing periods. Using $T-\Tamb$
makes zero initialization consistent with thermal equilibrium
and avoids the singular $1/T$ evaluation associated with
initializing an absolute-temperature node at $0$~K.
A weak connection to $x_0$ regularizes the otherwise singular
zero-current operating-point equation and selects $x=x_0$.
Its transient relaxation rate, $g_x=10^{-9}$~s$^{-1}$,
displaces $x$ by $2\times10^{-7}$ over the five-period
test bench.}

\new{OpenVAF compiles the model to a shared object loaded
through the OSDI interface of \texttt{ngspice}~47.
Two model cards on independent nodes, driven by the same
current source at $\Ea=0.7$~eV and $\Ea=0$, produce the
activated and isothermal responses in one run. Rather than
reading the internal state, it is recovered from the terminal
quantities as $x=(v/i-\Roff)/(\Ron-\Roff)$ for $i\ne0$.}

\new{\Cref{tab:veriloga} compares the settled-period results
with the reference implementation of \cref{sec:numresults}.
The largest voltage discrepancy across eight cases is
$7\,\mu\mathrm{V}$ on a $79$~mV peak, consistent with
differences between the integration engines.
At $\Ea=0$, the classical isothermal excursion is recovered;
peak temperature rises agree with the adaptive reference
within $0.02$~K. The corresponding $v$--$i$ loops in
\cref{fig:veriloga} document this recovery and the agreement
between implementations.}

\begin{table}[ht!]
\centering
\caption{\new{ATDM compiled with OpenVAF and executed in \texttt{ngspice}~47,
over the settled forcing period at $f=0.015$~Hz, against the reference
implementation. $\Delta T$ is the peak rise above ambient; $\delta v$ is the
largest deviation from the reference, as a percentage of the peak voltage.}}
\label{tab:veriloga}
\new{\small\begin{tabular}{@{}lcrr@{}}
\toprule
$I_0$      & $\Delta T$ (K) & \multicolumn{2}{c}{$\delta v$ (\% of peak)} \\
\cmidrule(l){3-4}
($\mu$A)   & model / ref.   & $\Ea=0.7$ & $\Ea=0$ \\
\midrule
$2$ & $2.17$ / $2.16$   & $0.0004$ & $0.0004$ \\
$3$ & $5.64$ / $5.63$   & $0.0006$ & $0.0005$ \\
$4$ & $13.22$ / $13.20$ & $0.0015$ & $0.0007$ \\
$5$ & $24.79$ / $24.79$ & $0.0088$ & $0.0007$ \\
\bottomrule
\end{tabular}}
\end{table}

\new{The model source (\texttt{atdm.va}), the four test benches, the scripts
that compile and run them on Linux and on Windows, and the script that
compares their output with the reference implementation are available from
the corresponding author upon reasonable request. A further diagnostic bench
perturbs each model-card parameter in turn to check that it reaches the
compiled model.}

\end{appendices}

{\footnotesize
\bibliographystyle{snstyle}
\bibliography{Bibliographie}
}

\end{document}